\documentclass[10pt, conference, compsocconf]{IEEEtran}
\usepackage{latexsym}
\usepackage{times}
\usepackage{graphicx}
\usepackage{subfigure}
\usepackage{multirow}
\usepackage{amsfonts}
\usepackage{amsmath}
\usepackage{caption}
\usepackage{algorithm,algorithmic}
\usepackage[usenames]{color}
\usepackage{url}

\newcommand{\comment}[1]{}

\newsavebox{\boxone}
\newsavebox{\boxtwo}
\newsavebox{\boxthree}

\newlength{\narrow}
\newlength{\cnarrow}
\newcommand{\topline}{
  \hrule
  \vskip .5\baselineskip}
\newcommand{\bottomline}{
  \vskip 2pt
  \hrule}

\newcommand{\chbox}[2]{
  \hbox to #1{\hss\vtop{#2}\hss}}

\newcommand{\nchbox}[1]{
  \chbox{\narrow}{#1}}

\newcommand{\cnchbox}[1]{
  \chbox{\cnarrow}{#1}}

\newcommand{\fcode}[1]{
  
  \chbox{\textwidth}{\tgrind\input{#1}}}

\newcommand{\ncode}[1]{
  
  \chbox{\narrow}{\tgrind\input{#1}}}

\def\nfig#1#2#3{
  \vtop{\nchbox{#1}
  \hbox to\narrow{\parbox{\narrow}{\caption{#2}\label{#3}}}}}

\newcommand{\cncode}[1]{
  \chbox{\cnarrow}{\tgrind\input{#1}}}

\def\codefiggen[#1]#2#3#4#5#6{
  \begin{figure}[#1]
  #5
  \fcode{#2}
  \center\parbox{.9\textwidth}{\caption{#3}\label{#4}}
  #6
  \end{figure}}

\def\codefig[#1]#2#3#4{
  \codefiggen[#1]{#2}{#3}{#4}{}{}}

\def\codefigline[#1]#2#3#4{
  \codefiggen[#1]{#2}{#3}{#4}{\topline}{\bottomline}}

\def\doublefiggen[#1]#2#3#4#5#6#7#8#9{
  \begin{figure}[#1]
  #8
  \hbox to \textwidth{
  \nfig{#2}{#3}{#4}
  \hfil
  \nfig{#5}{#6}{#7}}
  #9
  \end{figure}}

\def\doublefig[#1]#2#3#4#5#6#7{
  \doublefiggen[#1]{#2}{#3}{#4}{#5}{#6}{#7}{}{}}

\def\doublefigline[#1]#2#3#4#5#6#7{
  \doublefiggen[#1]{#2}{#3}{#4}{#5}{#6}{#7}{\topline}{\bottomline}}

\def\doublecodefig[#1]#2#3#4#5#6#7{
  \doublefig[#1]{\ncode{#2}}{#3}{#4}{\ncode{#5}}{#6}{#7}}

\def\doublecodefigline[#1]#2#3#4#5#6#7{
  \doublefigline[#1]{\ncode{#2}}{#3}{#4}{\ncode{#5}}{#6}{#7}}

\newcommand{\codepair}[4]{\vbox{
  \hbox{\ncode{#1} \hfil \ncode{#3}}
  \vskip .3\baselineskip plus .3\baselineskip
  \hbox{\hbox to\narrow{#2\hfil} \hfil \hbox to\narrow{#4\hfil}}}}

\def\codepairfig[#1]#2#3#4#5#6#7{
  \begin{figure}[#1]
  \codepair{#2}{#3}{#4}{#5}
  \center\parbox{.9\textwidth}{\caption{#6}}
  \label{#7}
  \end{figure}}

\def\cncodepairfiggen[#1]#2#3#4#5#6#7{
  \begin{figure}[#1]
  #6
  \hbox{\cncode{#2}\hfil\cncode{#3}}
  \center\parbox{.9\columnwidth}{\caption{#4}\label{#5}}
  #7
  \end{figure}}

\def\cncodepairfig[#1]#2#3#4#5{
  \cncodepairfiggen[#1]{#2}{#3}{#4}{#5}{}{}}

\def\cncodepairfigline[#1]#2#3#4#5{
  \cncodepairfiggen[#1]{#2}{#3}{#4}{#5}{\topline}{\bottomline}}

\def\doublefigOnecap*[#1]#2#3#4#5{
  \begin{figure*}[#1]
  \hbox to \textwidth{
  \nchbox{#2}
  \hfil
  \nchbox{#3}}
  \caption{#4}
  \label{#5}
  \end{figure*}}

\def\doublefigOnecap[#1]#2#3#4#5{
  \begin{figure}[#1]
  \topline
  \hbox to \columnwidth{
  \cnchbox{#2}
  \hfil
  \cnchbox{#3}}
  \caption{#4}
  \label{#5}
  \bottomline
  \end{figure}}

\def\PSfig[#1]#2#3#4{
 \begin{figure}
 \centerline{\psfig{file=#2,width=\columnwidth}}
 \caption{{#3}}
 \label{#4}
 \end{figure}}

\def\PSfiglines[#1]#2#3#4{
 \begin{figure}[#1]
 \topline
 \centerline{\psfig{file=#2,width=\columnwidth}}
 \caption{{#3}}
 \label{#4}
 \bottomline
 \end{figure}}

\def\PSfiglinesht[#1]#2#3#4#5{
 \begin{figure}[#1]
 \topline
 \centerline{\psfig{file=#2,height=#3}}
 \caption{{#4}}
 \label{#5}
 \bottomline
 \end{figure}}

\def\doublePSfig[#1]#2#3#4#5#6{
  \doublefigOnecap[#1]
    {\cnchbox{\psfig{file=#2,height=#4}}}
    {\cnchbox{\psfig{file=#3,height=#4}}}
    {#5}
    {#6}}

\newlength{\boxwidth}
\newcommand{\bproof}{{\bf Proof Sketch:}}
\newcommand{\eproof}{\mbox{$\Box$}}

\def\tabdoublecode#1#2#3#4{
 \begin{figure*}[t]
 \topline\vs{-.4}
 \hbox to \columnwidth{
 \vtop{\small
 \begin{tabbing}
 #1
 \end{tabbing}}
 \hfil
 \hfil
 \hfil
 \vtop{\small
 \begin{tabbing}
 #2
 \end{tabbing}}
 }
 \caption{#3\label{#4}}
 \bottomline
 \end{figure*}
}
\def\tabtriplecode#1#2#3#4#5{
 \begin{figure}
 \topline\vs{-.4}
 \hbox to \columnwidth{
 \vtop{\small
 \begin{tabbing}
 #1
 \end{tabbing}}
 \hfil
 \hfil
 \hfil
 \vtop{\small
 \begin{tabbing}
 #2
 \end{tabbing}}
 \hfil
 \hfil
 \hfil
 \vtop{\small
 \begin{tabbing}
 #3
 \end{tabbing}}
 }
 \caption{#4\label{#5}}
 \bottomline
 \end{figure}
}

\newtheorem{lemma}{Lemma}
\newcommand{\blemma}{\begin{lemma}}
\newcommand{\elemma}{\end{lemma}}

\newtheorem{thm}{Theorem}
\newcommand{\bthm}{\begin{thm}}
\newcommand{\ethm}{\end{thm}}

\newtheorem{defin}{Definition}
\newcommand{\bdefin}{\begin{defin}}
\newcommand{\edefin}{\end{defin}}

\newtheorem{observ}{Observation}
\newcommand{\bobserv}{\begin{observ}}
\newcommand{\eobserv}{\end{observ}}

\newcommand{\vs}[1]{\vspace{#1cm}}
\newcommand{\be}{\begin{equation}}
\newcommand{\ee}{\end{equation}}

\newcommand{\bdesc}{\begin{description}}
\newcommand{\edesc}{\end{description}}
\newcommand{\benum}{\begin{enumerate}}
\newcommand{\eenum}{\end{enumerate}}
\newcommand{\bitem}{\begin{itemize}}
\newcommand{\eitem}{\end{itemize}}
\newcommand{\bcenter}{\begin{center}}
\newcommand{\ecenter}{\end{center}}
\newcommand{\btabular}{\begin{tabular}}
\newcommand{\etabular}{\end{tabular}}
\newcommand{\beqnarr}{
 \begin{eqnarray}}
\newcommand{\eeqnarr}{\end{eqnarray}}

\begin{document}

\title{Distance Preserving Graph Simplification}




\author{
    Ning Ruan \\
    Department of Computer Science \\
    Kent State University \\
    nruan@cs.kent.edu
\and
    Ruoming Jin \\
    Department of Computer Science \\
    Kent State University \\
    jin@cs.kent.edu
\and
    Yan Huang \\
    Department of CSE \\
    University of North Texas \\
    huangyan@unt.edu
}


\maketitle

\begin{abstract}
Large graphs are difficult to represent, visualize, and understand. In this paper, we introduce ``gate graph'' - a new
approach to perform graph simplification. A gate graph provides a simplified topological view of the original graph.
Specifically, we construct a gate graph from a large graph so that for any ``non-local'' vertex pair (distance higher
than some threshold) in the original graph, their shortest-path distance can be recovered by consecutive ``local''
walks through the gate vertices in the gate graph. We perform a theoretical investigation on the gate-vertex set
discovery problem. We characterize its computational complexity and reveal the upper bound of minimum gate-vertex set
using VC-dimension theory. We propose an efficient mining algorithm to discover a gate-vertex set with guaranteed
logarithmic bound. We further present a fast technique for  pruning redundant edges in a gate graph. The detailed
experimental results using both real and synthetic graphs demonstrate the effectiveness and efficiency of our approach.

\end{abstract}

\section{Introduction}
\label{intro}

\comment{
Graphs have been widely used as a fundamental abstraction to model and analyze data from a variety of real-world applications, ranging from biological, ecological, and mobile caller systems, to the World Wide Web, social networks, and financial markets.
In these applications, graphs provide an effective mechanism to represent the interrelationships between different entities in the system.}

Reducing graph complexity or graph simplification is becoming an increasingly important research topic~\cite{AggarwalWang10,SpielmanS2008,Zhou10,ToivonenMZ10,DBLP:conf/sigmod/SatuluriPR11}.
It can be very challenging to grasp a graph even with thousands of vertices.
Graph simplification targets at reducing edges, vertices, or extracting a high level abstraction of the original graph so that the overall complexity of the graph is lowered while certain essential properties of the graph can still be maintained.
It has been shown that such simplification can help understand the underlying structure of the graph~\cite{Hennessey:2008,AggarwalWang10}; better visualize graph topology~\cite{DBLP:conf/visualization/RafieiC05,Hennessey:2008};
and speed up graph computations~\cite{Biedl:2000,Misiolek:2006,Feder:1991,Karande:2009:SUA,Tao11,DBLP:conf/sigmod/SatuluriPR11}.

\comment{
Graphs are becoming very large at a faster pace than ever. For example, facebook has 500 million active users interacting with 900 million objects \cite{facebookurl}. It is very challenge to grasp a graph with a large number of vertices and edges and graph simplification is becoming an increasingly important research topic. }

\comment{
Though graph clustering~\cite{AggarwalWang10} or decomposition~\cite{qiu:spectral} can provide a high-level view of a graph, clusters are generally too coarse to represent the essential information of the graph (such as the connectivity and shortest-path distance measure).
Recently, several works study how to simplify the graphs while maintaining their key graph properties, such as all-pair effective resistance~\cite{SpielmanS2008}, connectivity~\cite{Zhou10}, and other path-oriented  measures~\cite{ToivonenMZ10}.
Indeed, in these studies, the simplified graph is a spanning subgraph which is produced by removing non-essential edges from the original graphs without affecting the desired goodness function measuring the corresponding graph property.}

In this paper, we investigate how to extract a set of vertices from a graph such that the vertex locations and relationships not only help to preserve the distance measure of the original graph, but also provide a simplified topological view of the entire graph. Intuitively, these vertices can be considered to distribute rather ``evenly'' in the graphs in order to reflect its overall topological structure. For any ``non-local'' vertex pair (distance higher than some threshold), their shortest-path distance can be recovered by consecutive ``local'' walks through these vertices.
Basically, these vertices can be viewed as the key intermediate highlights of the long-range (shortest-path distance) connections in the entire graph.
In other words, for any vertex to travel to another vertex beyond its local range,
it can always use a sequence of these discovered vertices (each one being in the local range of its predecessor)
to recover its shortest path distance to the destination.
Thus, conceptually, this set of vertices form a ``wrap'' surrounding any vertex in the original graph,
so that any long range (shortest-path) traffic goes through the ``wrap''.
From this perspective, these vertices are referred to as the {\em gate} vertices and our problem is referred to as the {\em \textcolor{black}{gate-vertex set} discovery} problem.
Furthermore, these gate vertices can be connected together using only ``local'' links to form a {\em gate graph}. A gate graph not only reveals the underlying highway structure, but also can serve as a simplified view of the entire graph.
Gate-vertex set and gate graph have many applications in graph visualization ~\cite{HermanGuyS00, deSilvTenenbaum03} and shortest path distance computation~\cite{Tao11}.

\subsection{Problem Definition}
\label{problem}
Let $G=(V,E)$ be an unweighted and undirected graph, where $V=\{1,2,..., N\}$ is the vertex set and $E \subseteq V \times V$ is the edge set of graph $G$.
We use $(u,v)$ to denote the edge from vertex $u$ to vertex $v$, and $P_{v_0,v_p}=(v_0,v_1,...,v_p)$ to denote a simple path from vertex $v_0$ to vertex $v_p$.
The length of a simple path in unweighted graph is the number of edges in the path.
For weighted graph, each edge $e \in E$ is assigned  a weight $w(e)$.
The length of a simple path in a weighted graph is the sum of weights from each edge in the path.
The distance from vertex $u$ to vertex $v$ in the graph $G$ is denoted as $d(u,v)$, which is the minimal length of all paths from $u$ to $v$.



Given a user-defined threshold $\epsilon>0$, for any pair of {\em connected} vertices $u$ and $v$, if their distance is strictly less than $\epsilon$ ($d(u,v) < \epsilon$), we refer to them as a {\bf local pair}, and their distance is referred to as a {\bf local distance}; if their distance is higher than or equal to $\epsilon$ but finite, we refer to them as a {\bf non-local pair}, and their distance is referred to as a {\bf non-local distance}.
In addition, we also refer to $\epsilon$ as the {\em locality} parameter or the {\em granularity} parameter.

\bdefin({\bf \textcolor{black}{Minimum Gate-Vertex} Set Discovery (MGS) Problem})
\label{gatevertexset}
Given an unweighted and undirected graph $G=(V,E)$ and user-defined threshold $\epsilon>0$, vertex set $V^\ast \subseteq V$ is called a gate-vertex set if $V^\ast$ satisfies the following property: for any {\bf non-local pair} $u$ and $v$ ($d(u,v) \ge \epsilon$),
there is a vertex sequence formed by consecutive {\bf local pairs} from $u$ to $v$, $(u,v_1,v_2, \cdots, v_k,v)$ where $v_1, v_2, \cdots, v_k \in V^\ast$, such that $d(u,v_1)< \epsilon$, $d(v_1,v_2) < \epsilon$, $\cdots$ , $d(v_k,v) < \epsilon$, and $d(u,v_1)+d(v_1,v_2)+\cdots+d(v_k,v)=d(u,v)$. The gate vertex \textcolor{black}{set} discovery problem seeks a set of gate vertices with smallest cardinality.
\edefin

In other words, the gate-vertex set guarantees that the distance between any {\em non-local pair} $u$ and $v$ can be recovered  using the distances from source vertex $u$ to a gate vertex $v_1$, between consecutive gate vertices, and from the last gate vertex $v_k$ to the destination vertex $v$. These are all local distances.
Here, the local distance requirement for recovering any non-local distance enables the gate vertices to reflect enough details of the underlying topology of the original graph $G$.
Based on the gate-vertex set, we can further define the {\em gate graph}.

\bdefin{\bf (Minimum Gate Graph Discovery (MGG) Problem)}
\label{gsprobdef}
Given an \textcolor{black}{unweighted} and undirected graph $G=(V,E)$ and a gate-vertex set $V^\ast$ ($V^\ast \subseteq V$) with respect to parameter $\epsilon$, the gate graph $G^\ast=(V^\ast,E^\ast,W)$ is any weighted and undirected graph where $W$ assign each edge $e \in E^\ast$ a weight $w(e)$, such that for any non-local pair $u$ and $v$ in $G$ ($d(u,v) \ge \epsilon$), we have $d(u,v)=$
{\small
\beqnarr
 min_{d(u,x) < \epsilon \wedge d(y,v) < \epsilon \wedge x,y \in V^\ast}
              d(u,x)+d(x,y|G^\ast)+d(y,v); \nonumber
\eeqnarr}
Here $d(x,y|G^\ast)$ is the distance between $x$ and $y$ in the {\em weighted} gate graph.
The gate graph discovery problem seeks the gate graph with the minimum number of edges.
Note that the edges in the gate graph may not belong to the original graph.
\edefin

\comment{
 Here we illustrate two:

\noindent{\bf Graph Visualization:}
Graph visualization provides an human-centric mechanism to investigate graph characteristics and  to discover hidden patterns.
Though quite a few visualization approaches have been developed for graphs~\cite{HermanGuyS00}, how to handle large graphs and how the visualization can be more informative with respect to graph properties, such as distance and connectivity, are still two key challenges facing graph visualization.
The discovered gate-vertex set and gate graph can help scale the existing visualization approaches by allow them to focus on the positions of a small number of key vertices and their relationships, which still providing good approximation  of the overall graph topology.
In particular, the discovered gate-vertices can serve as the landmarks to scale the popular multidimensional scaling methods (MDS)~\cite{deSilvTenenbaum03}.
The relationship between gate vertices and landmarks are further discussed in Section~\ref{related}.
In addition, the gate-vertex set and gate graph provide useful and informative views for understanding the shortest-path distance related graph properties  due to their inherent characteristics.

\noindent{\bf Distance Computation:}
Computing shortest-path distance is a fundamental task in graph mining and management.
In addition, many important graph properties, such as graph diameter, betweenness centrality and closeness centrality, are all highly dependent on the distance computation.
Even though the BFS approach for computing pair-wise distance is quite efficient for small graphs, it can be quite expensive for large graphs.
Leveraging the highway structure to speed up the distance computation has been shown to be quite successful in road-network and planar graphs~\cite{Jing98,Jung02,Sanders05}.
However, how to construct a similar structure in the general graph is still an open problem.
Here, the  gate graph provides a natural solution to this problem.
}

\comment{
Furthermore, using the gate graph, we can explicitly decompose the distance computation  into the local distance computation ($d(u,x), d(y,v)$) and the gate distance computation ($d(x,y)$).
This can be very useful in distance computation in the emerging cloud computing environment.
Due to privacy concerns, a user may not want to load the entire graph which would place much sensitive information in the cloud-server side.
By only publishing the gate graph, the user is able to effectively utilize the cloud environment without disclosing sensitive information.
In particular, the parameter $\epsilon$ can help to balance the tradeoff between privacy and computational efficiency in such scenario.
}

\comment{
\noindent{\bf Graph Characterization and Comparison:}
Given a graph $G$, the proportion gate vertices out of all vertices in $G$ is a function  of $\epsilon$.
When the $\epsilon$ increases, the number of gate vertices reduces.
For two different graphs, this function is also expected to be different.
However, when they are similar to one another, this function can also reflect that.
Thus, this function can be used to characterize a graph (being complementary to graph diameters, average shortest-path distance, etc.) and even to compare graphs.
In addition, since superimposing two large graphs can be very expensive~\cite{RiesenJB10},
the gate graph can be utilized to provide approximate matching at the global level. }

\noindent{\bf Our Contributions: }

\noindent 1) We introduce and formally define the new gate-vertex set and gate graph discovery problems, which are applicable to numerous graph mining tasks;

\noindent 2) Based on basic properties of gate vertices, we perform a theoretical study on gate-vertex set by connecting it to the theory of VC-dimension, and prove \textcolor{black}{NP-hardness of minimum gate-vertex set discovery problem};

\noindent 3) We develop an efficient mining algorithm based on the set-cover framework to discover the gate-vertex set with guaranteed logarithmic approximation bound.
We discuss a fast approach to prune redundant edges in gate graph;


\noindent 4) We perform a detailed experimental evaluation using both real and synthetic graphs. Our results demonstrate the effectiveness and efficiency of our approach.

\section{Related Problems and Work}
\label{related}

\comment{
We first review related works on {\em graph simplification} and {\em shortest-path distance computation}. We then review related graph concepts, {\em landmarks} and {\em graph separators}, and compare them to gate vertices.
}
\noindent{\bf Graph Simplification:}
Our work on discovering a gate-vertex set and gate graph can be categorized as graph simplification with focus on preserving shortest path distance measure.
The most intuitive graph simplification method is graph clustering~\cite{AggarwalWang10} or decomposition~\cite{qiu:spectral}, which provides a high-level view of the graphs. However, this approach mainly focuses on discovering the community structure of the graph, and its representation is generally too coarse to preserve many other essential information of the graphs (such as the connectivity and shortest-path distance measure).
Several recent efforts study how to simplify the graphs while maintaining its key graph properties, such as the effective resistance~\cite{SpielmanS2008}, connectivity~\cite{Zhou10}, and other path-oriented measures~\cite{ToivonenMZ10}.
However, in these studies, the simplified graph is a spanning subgraph of the original graph, and thus does not reduce the overall scale of the graph in terms of the number of vertices.
In our work, we instead focus on discovering a subset of essential vertices which can maximally recover the all-pair shortest-path distances with respect to the locality parameter $\epsilon$.

In order to better visualize a large graph, the visualization community has proposed several methods to simplify graphs.
For instance, authors in ~\cite{DBLP:conf/visualization/RafieiC05} consider sampling a subgraph from the original graph for visualization, and in \cite{Hennessey:2008}, the authors develop a pruning framework to remove unimportant vertices in terms of their betweenness and other distance-related measures.
These methods are in general heuristically-oriented and cannot provide quantitative guarantee on how good the graph properties are preserved.
Our gate-vertex set and gate graph provide a new means to visualize large graphs and assist distance-centered graph visualization and analysis.

\comment{
In addition, several studies focus on simplifying graphs in order to speed up the computational tasks in graphs.
In ~\cite{Biedl:2000,Misiolek:2005,Misiolek:2006}, edge (and vertex) reduction techniques are studied to improve the computational efficiency of the network flow problem.
More recently~\cite{Karande:2009:SUA}, cliques or dense subgraphs are detected and compressed into virtual vertices to speed up the all-pair shortest-path computation, edge and vertex connectivity, matrix multiplication, PageRank, etc.
From this perspective, the gate graph can help facilitate the shortest path distance computation.}

Finally, several works~\cite{DBLP:conf/kdd/FaloutsosMT04,DBLP:conf/kdd/TongF06,DBLP:journals/datamine/HintsanenT08,DBLP:conf/pakdd/KasariTH10} study how to extract a concise subgraph which can best describe the relationship between a pair  or a set of vertices in terms of electric conductance~\cite{DBLP:conf/kdd/FaloutsosMT04, DBLP:conf/kdd/TongF06} or network reliability~\cite{DBLP:journals/datamine/HintsanenT08,DBLP:conf/pakdd/KasariTH10}. Our goal is to depict the \textcolor{black}{shortest-path} distances using gate vertices and gate graph.

\comment{
\begin{figure}
\centering
\begin{tabular}{c}
\psfig{figure=Figures/kskipsp.eps,scale=0.4}
\end {tabular}
\caption {Highlighted $4$-skip cover shortest path}
\label{4skipcover}
\end{figure}
}

\noindent{\bf Shortest-Path Distance Computation:}
Computing shortest-path distance is a fundamental task in graph mining and management.
Many important graph properties, such as graph diameter, betweenness centrality and closeness centrality, are all highly dependent on distance computation.
Even though the BFS approach for computing pair-wise distance is quite efficient for small graphs, it is very expensive for large graphs. Leveraging the highway structure to speed up the distance computation has been shown to be quite successful in road-network and planar graphs~\cite{Jing98,Jung02,Sanders05,Tao11}.
The recent $k$-skip graph \cite{Tao11} work represents the latest effort in using highway structure to reduce the search space of the well-known shortest distance computation method, {\em Reach} \cite{gutman:reach-based}.
Basically, each shortest path is succinctly represented by a subset of vertices, namely $k$-skip shortest path, such that it should contain at least one vertex out of every $k$ consecutive vertices in the original shortest path.
In other words, $k$-skip shortest path compactly describes original shortest path by sampling its vertices with a rate of at least $1/k$.
Tao {\em et al.} show that those sampled vertices can be utilized to speed up the distance computation.
Following the similar spirit, gate-vertex set and gate graph directly highlight the long-range connection between vertices, and can also serve as a highway structure in the general graph.

We note that the $k$-skip cover and gate vertices are conceptually close but different.
The $k$-skip \textcolor{black}{cover} intends to uniformly sample vertices in shortest paths, whereas the gate-vertex \textcolor{black}{set} tries to recover shortest-path distance using intermediary vertices (and local walks).
More importantly, the $k$-skip cover focuses on the road network and implicitly assume {\em there is only one shortest path between any pair of vertices}~\cite{Tao11}.
In this work, we study the generalized graph topology, where there may exist more than one shortest path between two vertices which is very common in graphs such as a social network. Our goal is to recover the non-local distance between any pair of vertices using only one shortest path.
Finally, in this paper, we focus on developing methods to discover minimum gate-vertex set, whereas \cite{Tao11} only targets at a random set of vertices which forms a $k$-skip cover, i.e., the minimum $k$-skip cover problem is not addressed.


\comment{
As shown in Figure \ref{4skipcover}, $4$-skip shortest path is highlighted, such that there is at least one vertex of every $4$ consecutive vertices is included.
All $k$-skip shortest path constitutes $k$-skip cover,
and $k$-skip graph consists of $k$-skip cover and essential connections among them (i.e., in undirected graph, the distance between two vertices in $k$-skip gate is no greater than $k$ ).
Utilizing $k$-skip graph with smaller number of vertices and edges, the significant performance improvement on computing shortest distance is possible to be gained over that on original graph.
Following the similar spirit, the discovered gate-vertex set and gate graph directly highlight the long-range connection between vertices, and become a natural candidate for the highway structure in the general situation.
However, as mentioned in \cite{Tao11}, {\em how to design a deterministic algorithm for finding minimum $k$-skip cover} is still open problem.
Also, the complexity of {\em minimum $k$-skip cover problem} is unexplored.
We will discuss discuss the relationship between gate-vertex set discovery problem and minimum $k$-skip cover problem later,
and resolve aforementioned concerns by extending our theoretical results and algorithmic solution.}

\comment{
Our work is also related to the $2$-hop~\cite{cohen2hop} index scheme which can handle reachability and distance queries in a unified framework.
In $2$-hop, each vertex $v$ records a list of intermediate vertices $L_{out}(v)$ which it can reach, and a list of intermediate vertices $L_{in}(v)$ which can reach it.
The shortest distance between vertex $u$ and vertex $v$ can be computed by selecting the minimal value of the sum of distance from $u$ to intermediate vertex $x$ and distance from $x$ to $v$, where $x$ is the common vertex between $L_{out}(u)$ and $L_{in}(v)$.
As we see, $2$-hop must record at least one vertex (i.e., $x \in L_{out}(u) \cap L_{in}(v)$) lying in a shortest path between any pair of vertices.
Its goal is to minimize the labeling cost which is defined as the total number of labels each vertex records, while ours is to extract a small set of gate vertices and their relationship (gate graph) to facilitate the distance-centered graph simplification and analysis.
In addition, we note that $2$-hop has been generalized to $3$-hop~\cite{DBLP:conf/sigmod/JinXRF09} for reachability queries directed graphs, utilizing a chain decomposition to simulate the highway structure.
However, how to extend $3$-hop to distance queries is still an open problem; this work may help address that problem.

Finally, the graph spanner approach~\cite{KP92} tries to discover a spanning subgraph of the original graph which can approximate the all-pair shortest path distance with desired estimation accuracy, often referred to as the stretch factor.
Indeed, such methods can also viewed as a graph simplification with distance preserving property.
As discussed earlier, our problem is different as it focuses on selecting a subset of vertices which can help recover the exact shortest-path distance.
}

\noindent{\bf Landmarks:}
Landmark vertices (or simply landmarks) are a subset of vertices in the graph which are selected and utilized for graph navigation (particularly shortest-path distance computation)~\cite{FJJJRSZ01,Ng01predictinginternet,Kleinberg04,GoldbergSODA05,Rattigan:2006,Potamias09,SarmaSMR10} and transformation (multidimensional scaling)~\cite{deSilvTenenbaum03}.
Given a landmark set, each vertex in the graph can approximate its network ``position'' by its distances to each  landmark.
Thus, each vertex is directly mapped to a multidimensional space where each landmark corresponds  to a unique dimension.
In online shortest-path distance computation, landmarks have been used together with triangle inequality for pruning search space~\cite{GoldbergSODA05}; several studies directly utilize landmarks for distance estimation~\cite{Kleinberg04,Rattigan:2006,Potamias09,SarmaSMR10}.
However, the landmarks generally are not necessarily good representatives for highlighting the underlying topology of the entire graphs, while the gate vertices explicitly ensure any pair-wise distance can be recovered through user-defined granularity threshold.

\comment{
In addition, how to select landmarks is a difficult problem, and popular methods are either based on sampling or a greedy procedure~\cite{Potamias09}. There is typically no overall quantitative measure for determining the goodness of landmarks. A particularly interesting work ~\cite{Potamias09} introduces the \textsc{Landmarks-Cover} problem which tries to find a minimum number of points such that for any pair of vertices $u$ and $v$, there exists at least one landmark in a shortest path from $u$ to $v$. This guarantees that each estimation is the exact distance.
The problem turns out to be NP-hard and the set-cover framework can solve it.
However, it needs to compute the all-pair shortest-path distances, which is too expensive for large graphs.
As we will show, our gate-vertex set discovery problem can be also transformed into a set-cover framework, but our algorithm needs to compute only local distances which can be computed rather easily.
}

\noindent{\bf Vertex Separators:}
Vertex separators~\cite{RosenbergH2001:GSA}  are a set of vertices (denoted as $S$) in a graph $G$ which partition the entire vertex set $V$ into three sets, $A$, $S$ and $B$, where there are no edges between vertices in $A$ and $B$.
Using vertex separators, a graph can be decomposed recursively. This is often used as a basis for applying a divide-and-conquer approach to (hard) graph problems.
\comment{
To achieve that, the number of the vertex separators generally needs to be small, and the size of sets $A$ and $B$ should be reasonably balanced.
Finding optimal graph separators is NP-hard and in most of the graphs, there is no small separator.
In some special graph families, such as planar graphs, small separators exist and efficient algorithms can discover them~\cite{LiptonTarjan1979}.
Moreover, vertex separators have been extensively used in road-networks for constructing highway structures~\cite{Jing98,Jung02} and to construct labeling schemes for shortest-path distance computation~\cite{MozesS10}.
}
The gate vertices are different from separators as they do not have to explicitly partition the graphs.
In particular, if there are multiple non-local shortest paths between two vertices, the gate vertices will guarantee to recover  at least one of them.
Thus, the gate vertices in some sense relax the condition of vertex separators and thus allow us to recover the shortest-path distance even on general graphs.

\comment{
\noindent{\bf Landmarks: }
They show  this problem is NP-hard and point out that a set-cover framework can solve this problem.
Note that this problem has a very close relationship to the aforementioned $2$-hop labeling scheme.
In $2$-hop, only a subset of relevant landmarks is assigned to a given node.
Here, they assign the complete set of landmarks to each node.
However, unlike $2$-hop, this method employs a set of heuristics based on the node properties, such as degree or centrality, to select the landmarks, and landmarking does not aim to provide exact approximation.
Instead, they use the small number of landmarks to estimate the SPD.

In graph analysis, a set of {\em landmarks} is often selected and utilized in assisting of
The basic idea of those approaches is to use a subset of vertices as intermediate points to estimate point-to-point shortest distance.
Especially, a set of vertices are selected as landmark based on certain criteria and each vertex records the shortest distances to those landmarks.
When the distance query between two vertices is coming, we can quickly estimate it by concatenating the distance from source to landmark and the distance from landmark to destination.
The recent landmark-based method is proposed by Potamias et. al. \cite{Potamias09} focuses on investigating the problem of good landmark selection.
They have proved that the optimal landmark selection problem is NP-hard, such that there is at least one landmark lying in the shortest path of any pair of vertices.
Therefore, several heuristics are introduced to efficiently address this problem in the large graph.
Some other related works can be found in \cite{FJJJRSZ01,Ng01predictinginternet,Kleinberg04}.
Note that those works are related to our problem, but essentially are different from each other.
Unlike landmark-based approaches, our goal is to find a central structure instead of a set of vertices which captures shortest distances and serves as the intermediate hop on distance computation.
In addition, we provide exact distance computation which cannot be guaranteed in landmark approaches.
}


\section{Properties of Gate-Vertex Set and Problem Transformation}
\label{gatevertexprop}

Based on the definition of the gate-vertex set, to verify that a given set of vertices $V^\ast$ is a \textcolor{black}{gate-vertex} set, the na\"{i}ve approach is to explicitly verify that the distance between every {\bf non-local pair}  $(u,v)$ can be recovered through some sequence of consecutive {\bf local pairs}: ($u,v_1$), ($v_1,v_2$), $\cdots$, ($v_k,v$), where all intermediate vertices $v_1, v_2, \cdots, v_k \in V^\ast$.
Clearly, this can be expensive and difficult to directly apply to discover the minimum number of gate vertices.
In Subsection~\ref{localcond}, we first discuss an alternative (and much simplified) condition, which enables the discovery of \textcolor{black}{gate-vertex} set using only local distance, and reveal the NP-hardness of minimum gate-vertex \textcolor{black}{set} discovery problem.
In addition, we utilize the VC-dimension theory to bound the size of gate vertices.

\comment{
\subsection{Two-Side Condition and Set-Cover-with-Pair Solution}
\label{SCP}

We start with the following theorem which introduces a simplified condition to determine the gate-vertex set.

\bthm{\bf (Two-Side Condition)}
\label{equivalent}
Given an undirected graph $G=(V,E)$ and the parameter $\epsilon$, suppose there is a subset of vertices $V^\prime \subseteq V$ satisfying the following condition: for any {\bf non-local pair} $u$ and $v$, there is $x,y \in V^\prime$, such that $d(u,x) < \epsilon$,  $d(y,x) < \epsilon$, and
\beqnarr
 d(u,v)= d(u,x) + d(x,y) + d(y,v) \textnormal{    ($\star$)}.\nonumber
\eeqnarr
Then $V^\prime$ is a gate-vertex set $V^\ast$.
Furthermore, the smallest cardinality of set $V^\prime$ is equal to the smallest cardinality of gate-vertex set $V^\ast$:
$min |V^\prime|=min |V^\ast|$.
\ethm
\bproof
We will prove the condition ($\star$) is a sufficient ($\Rightarrow$) and necessary ($\Leftarrow$) condition for determining a gate-vertex set.

\noindent{\bf ($\Rightarrow$):} If $V^\prime$ has this property, then we can iteratively recover the shortest-path distance through the consecutive local pairs.
Based on the condition, for any non-local pair $u$ and $v$, we can discover $x_1 \neq u$ and $y_1 \neq v$ in $V^\prime$, such that $d(u,v)=d(u,x_1)+d(x_1,y_1)+d(y_1,v)$, where $d(u,x_1) < \epsilon$ and $d(y_1,v) < \epsilon$.
If $d(x_1,y_1) < \epsilon$, then we find the sequence.
If not, applying this condition for the non-local pair ($x_1,y_1$), we have $x_2 \neq x_1$ and $y_2 \neq y_1$ in $V^\prime$, such that
$d(x_1,y_1)=d(x_1,x_2)+d(x_2,y_2)+d(y2,y_1)$ where $d(x_1,x_2) < \epsilon$ and $d(y_2,y_1) < \epsilon$.
Clearly, if we can continue this process, eventually we have $d(x_i,y_i)< \epsilon$ (including the case where $x_i=y_i$) since the graph is finite.
Thus, for any non-local pair $u$ and $v$, we can recover a sequence $(u,x_1,x_2, \cdots x_i,y_i$, $\cdots y_2,y_1,v)$,
where each connective pair is a local pair.
Thus, if a set $V^\prime$ has the property ($\star$), then, it is a gate-vertex set.

\noindent{\bf ($\Leftarrow$):} For any gate-vertex set $V^\ast$, it must have the property $(\star)$. This can be immediately derived from the gate-vertex set definition.

From the sufficient inference, we have $min |V^\prime| \leq min |V^\ast|$ and from the necessary inference,
we have $min |V^\prime| \geq min|V^\ast|$.  Taken together, we have $min |V^\prime|=min |V^\ast|$.
\eproof

Based on Theorem~\ref{equivalent}, we introduce the following equivalent problem for discovering gate-vertex set.

\bdefin({\bf Minimal Gate-Vertex Set (MGS) Problem})
\label{gatevertexsetnew}
Given unweighted undirected graph $G=(V,E)$ and user-defined threshold $\epsilon$, we would like to discover a minimal set of vertices $V^\ast \subseteq V$ (gate-vertex sets) satisfying the following property: for any {\bf non-local pair} $u$ and $v$ ($d(u,v) \ge \epsilon$), there are vertices $x$ and $y$ in $V^\ast$, such that $d(u,x)+d(x,y)+d(y,v)=d(u,v)$, where $0<d(u,x)< \epsilon$ and $0<d(y,v) < \epsilon$.
\edefin

We conjecture the MGS problem is NP-hard. Based on this definition, we can see that MGS is closely related to the recently proposed set-cover-with-pair (SCP) problem~\cite{DBLP:conf/fsttcs/HassinS05}.

\bdefin{\bf (Set-Cover-with-Pairs (SCP) ~\cite{DBLP:conf/fsttcs/HassinS05})}
Let $U$ be the ground set and let $S = \{1, \ldots, M\}$ be a set of objects. For every $\{i,j\}\subseteq S$, let $\mathcal{C}(i, j)$ be the collection of elements in $U$ covered by the pair $\{i, j\}$. The objective of the set cover with pairs (SCP) problem is to find a subset $S^\prime \subseteq S$ with minimal cardinality such that
{\small \beqnarr \mathcal{C}(S^\prime) =\bigcup_{\{i,j\}\subseteq S^\prime} \mathcal{C}(i, j) = U \nonumber \eeqnarr}
\edefin

Given this, we can transform MGS to an instance of SCP as follows.
Given graph $G=V$ and parameter $\epsilon$, let $U$ include all non-local pairs $(u,v)$ ($d(u,v) \ge \epsilon$), and $S=V$.
Then for any vertex pair $x,y$ in $S$, the following set $\mathcal{C}(x,y)$ is considered to be covered by $\{x,y\}$:
\beqnarr
\mathcal{C}(x,y)=\{(u,v)| d(u,v) \ge \epsilon \wedge 0<d(u,x)< \epsilon \wedge 0<d(y,v)< \epsilon \} \nonumber
\eeqnarr
Clearly, the subset $S^\prime \subseteq S$ with minimal cardinality to cover all elements in $U$ is the minimal gate vertex set.
However, using this transformation to solve MGS has the following problems.
First, to perform such transformation, we basically need to compute the all-pair shortest-path distance ($O(|V|(|V|+|E|))$ time complexity) and to store the distance matrix ($O(|V|^2)$ space complexity). Both can be quite expensive.
Furthermore, since SCP is NP-hard and its best available (greedy) approximation algorithm~\cite{DBLP:conf/fsttcs/HassinS05} yields an $O(\sqrt{|U|\log |U|})$ approximation ratio.
In other words, the cardinality of vertex-gate set discovered by the greedy algorithm~\cite{DBLP:conf/fsttcs/HassinS05} is guaranteed no higher than a factor $O(|V| \sqrt{\log |V|})$ of the minimal cardinality (the number of non-local pairs $|U|$ can be close to  $\frac{|V|(|V|-1)}{2}$).
Clearly, such approximation bound is meaningless in this problem as the cardinality of minimal vertex-set is bounded  by $|V|$.
Given this, we can see that using SCP to solve MGS is not a feasible solution.
Can we derive an efficient algorithm without all-pair shortest-distance computation and delivering better approximate bound?
We provide a positive answer to this question in next subsection.
}

\subsection{Local Condition and Problem Reformulation}
\label{localcond}

In order to design a more efficient and feasible algorithm, we explore the properties of gate vertices and observe that gate-vertex set can be efficiently checked by a very simple condition.
Let $G=(V,E)$ be an unweighted and undirected graph. For any vertex $u \in V$, its {\bf $\epsilon$-neighbors}, denoted as $N_{\epsilon}(u)$ is a set of vertices such that their distances to $u$ is no greater than $\epsilon$,
i.e., $N_{\epsilon}(u)=\{v \in V| 0<d(u,v) \le \epsilon\}$.
Let $L$ be a set of vertices and $S = \{(u_0,v_0), ..., (u_k, v_k)\}$ be a set of vertex pairs in the graph $G$.
We say that $L$ {\bf covers} $S$ if for each vertex pair $(u_i,v_i)\textcolor{black}{\in S}$ there is at least one vertex $x \in L$ such that $d(u_i,v_i)=d(u_i,x) + d(x,v_i)$.

Now, we introduce the following key observation:


\comment{
\begin{figure}
\centering
\begin{tabular}{c}
\psfig{figure=Figures/localcond1.eps,width=3.0in,height=1.0in}
\end {tabular}
\caption {Proof of Lemma~\ref{restorelemma}}
\label{localcond1}
\vspace*{-3.0ex}
\end{figure}
}

\blemma{\bf (Sufficient Local Condition for MGS)}
\label{restorelemma}
If for each vertex $x$ in the graph $G$, there is a subset of vertices $L(x) \subseteq N_{\epsilon-1}(x)$ which covers all vertex pairs $\{(x,y_i) | d(x,y_i) = \epsilon \}$,
then $\bigcup_{x \in V} L(x)$ is a gate-vertex set of graph $G$.
In other words, a vertex set which covers any pair of vertices with distance $\epsilon$ is a gate-vertex set.
\elemma

\bproof
For any non-local pair $s$ and $t$ ($d(s,t) \ge \epsilon$), we denote one of their shortest paths to be
$P=(s=v_0,v_1,\cdots,v_{k}=t)$ with the length $k = d(s,t) \geq \epsilon$.
Let us consider $v_{\epsilon}$ on the shortest path.
Since $d(s,v_{\epsilon}) = \epsilon$, there is at least one vertex $x_0 \in L(s)$,
such that $d(s,x_0)+d(x_0,v_{\epsilon}) = d(s,v_{\epsilon})$.
Now, we consider two cases:

\noindent 1) If $d(x_0,t) < \epsilon$, then we recover a local-walk sequence ($s,x_0,t$).

\noindent 2) If $d(x_0,t) \ge \epsilon$, since $d(s,x_0)+d(x_0,t)=d(s,t)$ (based on the fact $d(x_0,v_{\epsilon})+d(v_{\epsilon},t)=d(x_0,t)$),
we have $d(x_0,t) < d(s,t)$.
Then, we can recursively apply the above method to identify $x_1$
between $x_0$ and $t$,
$x_2$ between $x_1$ and $t$, until $d(x_i,t) < \epsilon$.

Since $x_0, x_1, \cdots, x_i \in \bigcup_{x \in V} L(x)$ (i.e. they also belong to $V^\ast$) and the distance of every vertex pair $(x_m,x_{m+1})$ is less than $\epsilon$,
we can recover the distance between $s$ and $t$ to be $(s,x_0,x_1, \cdots, x_i, t)$, where $d(s,x_0)<\epsilon$, $d(x_0,x_1) < \epsilon$, $\cdots$, $d(x_i,t)<\epsilon$ and $d(s,x_0)+\sum_{m=0}^{i-1} d(x_m,x_{m+1}) +d(x_i,t)=d(s,t)$.
Therefore, $\bigcup_{x \in V} L(x)$ or any set of vertices $V^\prime$ which can cover any vertex pair with distance $\epsilon$, is a gate-vertex set.
\eproof

Interestingly, this local condition specified in Lemma~\ref{restorelemma} is also a necessary  one for a gate-vertex set.

\blemma{\bf (Necessary Local Condition for MGS)}
Given an \textcolor{black}{unweighted and undirected} graph $G$ and its gate-vertex set $V^\ast$ with respect to parameter $\epsilon$, for any vertex $s \in V$,
we have $L(s)=\{x \in V^\ast|0<d(s,x) < \epsilon\}$ such that for any vertex $t$ with distance $\epsilon$ to $s$ (i.e., $d(s,t)=\epsilon$),
there is $x \in L(s)$ with $d(s,t)=d(s,x)+d(x,t)$.
\elemma

\bproof
For any non-local vertex pair $(s,t)$ with $d(s,t) = \epsilon$, by the definition of gate-vertex set,
there must exist a sequence of vertices $x_0, x_1, ..., x_i \in V^\ast$, such that $d(s,t)=d(s,x_0)+d(x_0,x_1)+...+d(x_i,t)$ where $d(s,x_0)<\epsilon$, $\cdots$, $d(x_i,t)<\epsilon$.
Since $d(s,x_0) < \epsilon$, we have $x_0 \in L(s)$.
Also, it is easy to see that $d(s,t) = d(s,x_0)+d(x_0,t)$ because $d(s,t)-d(s,x_0)=d(x_0,x_1)+...+d(x_i,t) \ge d(x_0,t)$ and $d(s,t) \le d(s,x_0)+d(x_0,t)$.
Therefore, we have at least $x_0 \in L(s)$ satisfying $d(s,t) = d(s,x_0)+d(x_0,t)$.
\eproof

\comment{
\bproof
Following Theorem~\ref{equivalent},
Since $(u,v)$ is a non-local pair ($d(u,v) \ge \epsilon$), there must be vertices $x,y \in V^\ast$, such that $d(u,v)=d(u,x)+d(x,y)+d(y,v)$ ($0<d(u,x)< \epsilon$ and $0<d(y,v) < \epsilon$).
Clearly, it is also easy to see that $0<d(x,v)< \epsilon$ and $0<d(u,y) < \epsilon$.
Thus,  both $x$ and $y$ in $L(u)$ and can satisfy $d(u,v)=d(u,x)+d(x,v)$ and $d(u,v)=d(u,y)+d(y,v)$.
\eproof
}

Putting this together, given parameter $\epsilon$,
checking whether a subset of vertices $V^\ast \subseteq V$ is gate-vertex set is equivalent to checking the following condition: for any vertex pair $(u,v)$ with distance $\epsilon$,
there is a vertex $x \in V^\ast$ such that $d(u,v) = d(u,x)+d(x,v)$.
Similarly, we can rewrite the \textcolor{black}{minimum} gate-vertex set discovery problem in the following equivalent local condition (only covering \textcolor{black}{vertex} pairs with distance $\epsilon$).

\bdefin({\bf Minimum Gate-Vertex Set Problem using Local Condition}) \label{gatevertexsetlocal} Given unweighted
undirected graph $G=(V,E)$ and user-defined threshold $\epsilon$, we would like to seek a set of vertices $V^\ast$ with
minimum cardinality, such that any pair of vertices $(u,v)$ with distance $\epsilon$ is covered by at least one vertex
$x \in V^\ast$: $d(u,x)+d(x,v)=d(u,v)$. \edefin

In the following, we would like to prove NP-hardness of aforementioned problem by reducing the 3SAT problem.

\bthm {\bf (NP-hardness of MGS using Local Condition provided Shortest Paths)}
\label{nphardmgslocal}
Given a collection $P$ of vertex-pair $(u,v)$ with $d(u,v)=\epsilon$ denoting a set of shortest paths from unweighed undirected graph $G=(V,E)$,
finding minimum number of vertices $V^\ast \subseteq V$ such that any vertex-pair $(u,v)$ is covered by at least one vertex $x \in V^\ast$ is NP-hard.
\ethm

\bproof
We can reduce 3SAT problem to this problem.
Let $S$ be an instance of 3SAT with $n$ variables $x_1$, $x_2$, ..., $x_n$, and $m$ clauses $C_1$, $C_2$, ..., $C_m$.
We show that an instance of our problem can be constructed correspondingly as follows.
A unweighted undirected graph $G$ consisting of a vertex $p$ and a set of variable gadgets and clause gadgets will be generated.

The variable gadget with respect to variable $x$ contains $3$ vertices and $2$ edges:

\noindent 1) $3$ vertices: $b^x$, $b^{\overline{x}}$ and $e^x$;

\noindent 2) $2$ edges: $(b^x, e^x)$ and $(b^{\overline{x}}, e^x)$.

Also, we add edges $(p, b^x)$ and $(p, b^{\overline{x}})$ to build the connections between $p$ and variable $x$'s gadget.
For each clause $C_i=(X, Y, Z)$, we add vertex $c_i$ to graph $G$ first. Then, if $X=x$, we add edge $(b^x,c_i)$, otherwise, we add edge $(b^{\overline{x}}, c_i)$.
The same rule is applied to literals $Y$ and $Z$. Next we add 3 edges $(b^X, c_i)$, $(b^Y, c_i)$ and $(b^Z, c_i)$ into $G$.
The subgraph containing vertices $\{p,b^X,b^Y,b^Z,c_i\}$ and above created edges is called clause gadget regarding $C_i$.

Here we consider $\epsilon=2$ in the graph $G$.
That is, we try to find a set of vertices $V^\ast$ with minimum cardinality to cover a collection $P$ of shortest path with length $\epsilon$ (i.e., 2 in this scenario).
Note that, in our problem, for shortest path $SP=(x,y,z)$ with length $2$, only vertex $y$ in the middle can be used to cover $SP$ identified by its two endpoints $(x,z)$.
Let us define $P_{p e_i^x}$ to be vertex pairs indicating shortest paths with length $2$ between $p$ and $e^x$.
Moreover, $P_{p c_k}$ denotes vertex pairs representing shortest paths with length $2$ between $p$ and $c_k$.
We consider gate-vertex selection problem on vertex pairs $P= (\cup_i P_{p e^{x_i}}) \bigcup (\cup_k P_{p c_k})$ .

In the following, we prove that above 3SAT instance is satisfiable if and only if the instance of our problem has a solution of size at most $n$.
We need to prove both the ``only if'' and the ``if'' as follows.

\noindent $\Longrightarrow$: Suppose 3SAT instance $S$ is satisfiable and $f$ is its corresponding satisfying assignment.
For each variable $x$, if $f(x)=1$, vertex $b^x$ is added to $S^\ast$ to cover the shortest paths within $P_{p c_i}$ where clause $C_i$ contains $x$ or $\overline{x}$.
Otherwise, we add vertex $b^{\overline{x}}$ into $S^\ast$.
As we can see, either $b^x$ or $b^{\overline{x}}$ is selected, shortest paths within $P_{p e_i^x}$ always can be covered.
On the other hand, since one of literals $X$ in each clause $C$ is guaranteed to be true, there is always one vertex indicating such literal serves as intermediate hop in shortest paths of $P_{p c}$.
In this sense, only one vertex corresponding to the literal with true value is added to $V^\ast$ in each variable gadget.
Therefore, the solution size for the instance of our problem is $n$.

\noindent $\Longleftarrow$: Suppose graph $G$ has a gate-vertex set $V^\ast$ of size $n$ with respect to $\epsilon=1$.
For 3SAT instance $S$, we define a truth assignment by setting $f(x)=1$ if and only if vertex $b^x$ is included in $V^\ast$.
We will show this is satisfying assignment without conflict.
First, according to the definition of gate-vertex set, there are at least one vertex from $V^\ast$ cover vertex pair $P_{p c_i}$,
meaning that there are at least one literal with truth value existing in each clause.
This leads to the truth value of entire 3SAT instance.
Furthermore, in order to cover every vertex pair $P_{p e^{x_i}}$, either vertex $b^{x_i}$ or $b^{\overline{x_i}}$ must be chosen.
Considering the constraint $|V^\ast| \le n$, for each variable gadget, only one of $b^{x_i}$ and $b^{\overline{x_i}}$ can be included in gate-vertex set.
From the perspective of 3SAT instance $S$, this guarantees that only one of literals $x_i$ and $\overline{x_i}$ would be assigned with true value.
That is, no conflict occurs in our aforementioned assignment.
Putting both together, we can claim that $f$ is a satisfying truth assignment.
\eproof

\begin{figure}
\centering
\begin{tabular}{c}
\includegraphics[scale=0.6]{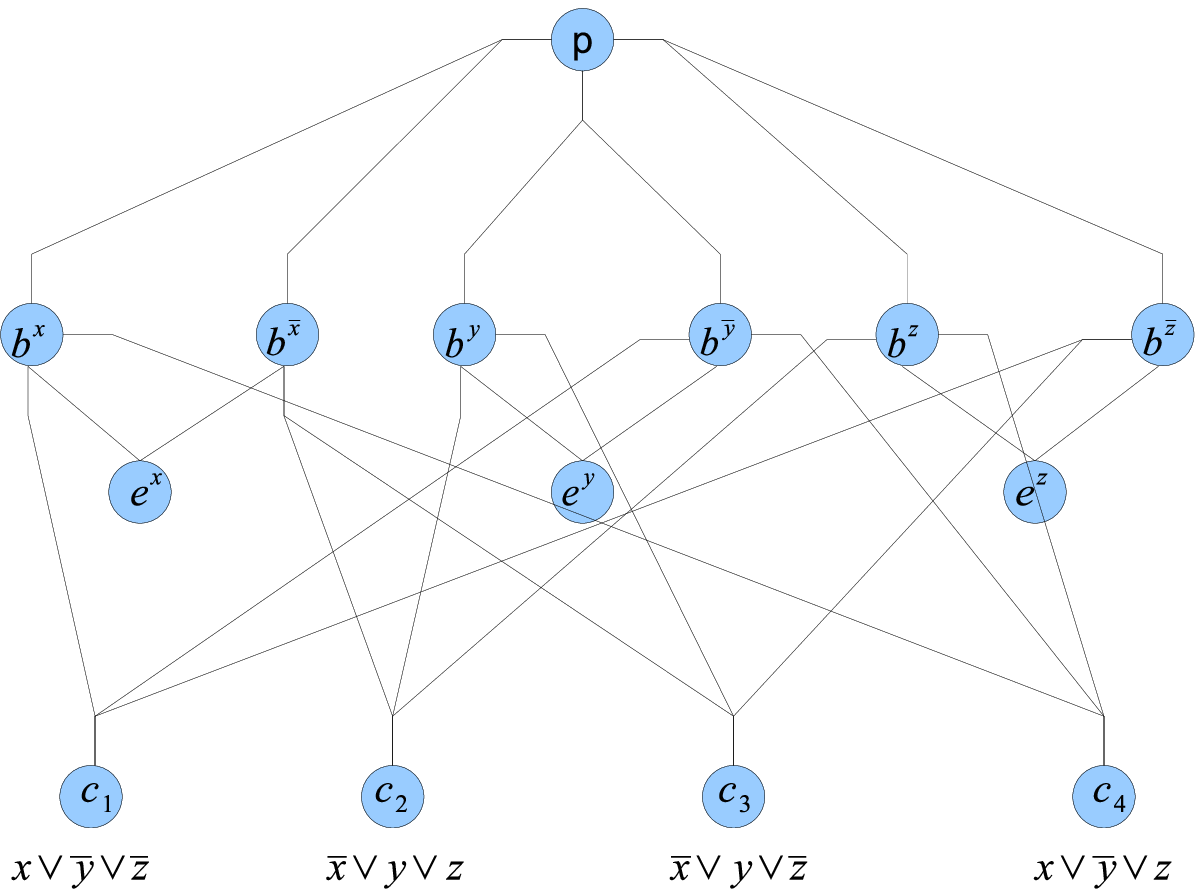}
\end {tabular}
\caption {Example for proof of Theorem \ref{nphardmgslocal}}
\label{nphardmgsexample}
\end{figure}

{\bf Example}: consider the following boolean formula $S$ in 3SAT problem with respect to variables $x$ $y$ and $z$:
\[ (x \vee \overline{y} \vee \overline{z}) \wedge (\overline{x} \vee y \vee z) \wedge (\overline{x} \vee y \vee \overline{z}) \wedge (x \vee \overline{y} \vee z) \]

To simplify the discussion, we name the above 4 clauses as $C_1$ $C_2$ $C_3$ and $C_3$, respectively.
Here, we consider the vertex pairs $P=\{(p,e^x),(p,e^y),(p,e^z), (p,c_1), (p,c_2), (p,c_3), (p,c_4)\}$ with distance $\epsilon$ (i.e., 2 in the example).
The constructed graph $G$ including variable gadgets and clause gadgets is shown in Figure~\ref{nphardmgsexample}.
In this example, the formula is satisfied by the assignment $x=1$, $y=1$ and $z=0$.
According to the rule defined in the proof, we add $b^x$, $b^y$ and $b^{\overline{z}}$ to gate-vertex set $V^\ast$.
It is not hard to verify that no vertex set with less vertices compared to $|V^\ast|$ can be obtained.
From another direction, we can see that $V^\ast = \{b^{\overline{x}}, b^{\overline{y}}, b^z \} \subseteq V(G)$ is minimum gate-vertex set with respect to vertex pairs in $P$
(i.e., the problem is equivalent to find the set cover problem: ground set $\{c_1,c_2,c_3,c_4\}$, candidate sets $\{b^X, b^Y, b^Z\}$ ($X \in\{x, \overline{x}\}$, $Y \in\{y, \overline{y}\}$ and $Z \in\{z, \overline{z}\}$)).
The corresponding satisfying assignment can be build as follows: $x=0$, $y=0$ and $z=1$.
It is straightforward to verify that this is a truth assignment for instance $S$.

\comment{
We prove Theorem~\ref{nphardmgslocal} by reducing the 3SAT problem to this problem. Due to the lack of space, the proof
is omitted here and can be found in the technical report~\cite{gategraphTR}.
}

\subsection{Size of Minimum Gate-Vertex Set}
\label{sizegateset}

\comment{
The performance improvement by using gate-vertex set highly depends on its size.
The question naturally arises: {\em can we derive a bound of minimum gate-vertex set which guarantees the performance gain? }
We will provide the positive answer in the following. }

In the following, using the theory of VC-dimension, we derive an upper bound of the cardinality of minimum gate-vertex set.

\noindent {\bf VC-dimension and $\epsilon$-net: } We start with a brief introduction of the VC-dimension of set systems
and $\epsilon$-net. The notion of VC-dimension originally introduced by Vladimir Vapnik and Alexey Chervonenkis in
\cite{VC1971} is widely used to measure the expressive power of a set system. Let $U$ be a finite set and $R$ a
collection of subsets of $U$, the pair $(U,R)$ is referred to be a set system. A set $A \subseteq U$ is {\em
shatterable} in $R$  if and only if for any subset $S$ of $A$, there is always a subset $X \in R$ where $X \cap A = S$.
In other words, $X$ contains the ``exact'' $S$ with no element in $A \backslash S$. Then, we say the VC-dimension of
set system $(U,R)$ is the largest integer $d$ such that no subset of $U$ with size $d+1$ can be shattered.
In addition, given parameter $\epsilon \in [0,1]$, a set $N \subseteq U$ is an $\epsilon$-net on $(U,R)$ if for any subset $X \in R$, $X$ has size no less than $\epsilon |U|$, the set $N$ contains at least one element of $X$.
For the set system with bounded VC-dimension $d$, the $\epsilon$-net theorem states there exists a $\epsilon$-net with size $O(\frac{d}{\epsilon} \log \frac{1}{\epsilon})$~\cite{Haussler:1986}.

Using the VC-dimension and $\epsilon$-net theorem, we can bound the size of minimum gate-vertex set.
\bthm
\label{minsizebound}
Given graph $G=(V,E)$ with parameter $\epsilon$, the size of minimum gate-vertex set is bounded by $O(\frac{|V|}{\epsilon-1} \log \frac{|V|}{\epsilon-1})$.
\ethm

To prove Theorem~\ref{minsizebound}, we need a few lemmas.
To facilitate our discussion, we introduce the following notations.
Given the input graph $G=(V,E)$, let $p^\ast_{s,t}$ be the subpath of shortest path $p_{s,t}$ without the two endpoints.
For instance, if $p_{s,t} = (s,u,...,v,t)$, its corresponding $p^\ast_{s,t}=(u,...,v)$.
Further, let $P_l$ only contains  shortest path $p_{s,t}$ of length $l$,
i.e., $P_l = \cup_{s,t} \{p_{s,t} \text{ s.t. } |p_{s,t}|=l \}$.
Given $P_l$ with $l \ge 1$, we say $P^\ast_l$ is a {\em core-set} of $P_l$ if for each shortest path $p_{s,t}$, only its subpath $p^\ast_{s,t}$ is included in $P^\ast_l$,
i.e., $P^\ast_l = \cup_{s,t} \{ p^\ast_{s,t} \text{ s.t. }  p_{s,t} \in P_l \}$.

We first establish the relationship between $\epsilon$-net and gate-vertex set.
\blemma {\bf ($\frac{\epsilon-1}{|V|}$-net)}
\label{gatesetnet}
Given a set system $(V,P^\ast_{\epsilon})$, where $P_{\epsilon}$ contains
a \textcolor{black}{shortest} path for every vertex pair with distance $\epsilon$ in graph $G=(V,E)$ ($P^\ast_{\epsilon}$ is the core-set of $P_\epsilon$), a $\frac{\epsilon-1}{|V|}$-net $V^\ast$ of $(V,P^\ast_{\epsilon})$ is a gate-vertex set.
\elemma
\bproof
By definition of $P^\ast_{\epsilon}$, the number of vertices in shortest path $p_{u,v} \in P^\ast_{\epsilon}$ is $\epsilon-1$.
According to definition of $\frac{\epsilon-1}{|V|}$-net, for each shortest path $p_{s,t} \in P^\ast_{\epsilon}$,
we have $p_{s,t} \cap V^\ast \neq \emptyset$.
Moreover, recall that each shortest path of $P^\ast_{\epsilon}$ is a subpath of some shortest paths of $P_{\epsilon}$ by removing two endpoints.
In other words, if $V^\ast$ contains at least one vertex from each shortest path in $P^\ast_{\epsilon}$,
then at least one vertex from shortest path in $P_{\epsilon}$ is included in $V^\ast$.
Since $P_{\epsilon}$ contains one shortest path for every vertex pair with distance $\epsilon$,
this satisfies the condition of gate-vertex set such that there is at least one vertex $x \in V^\ast$ holding $d(u,v)=d(u,x)+d(x,v)$ for every vertex pair $(u,v)$ with $d(u,v)=\epsilon$.
Therefore, $V^\ast$ is a gate-vertex set.
\eproof

To bound the size of $\epsilon$-net, the VC-dimension of the set system is needed.
In ~\cite{Goldberghd2,Tao11}, the VC-dimension of a {\em unique} shortest path system, i.e., only {\bf one} shortest path exists between any pair of vertices in a graph, is studied.
Formally, we first define {\em Unique Shortest Path System}:

\bdefin {\bf (Unique Shortest Path System (USPS))~\cite{Goldberghd2}}
Given a graph $G=(V,E)$ and a collection $Q$ of shortest paths from $G$,
we say $Q$ is a unique shortest path system if:
any vertex pair $u$ and $v$ is contained in two shortest \textcolor{black}{paths} $p_{s_1,t_1}, p_{s_2,t_2} \in Q$,
then they are linked by the same path, i.e., $p_{u,v}=p^\prime_{u,v}$,
where $p_{u,v}$ ($p^\prime_{u,v}$) is the subpath of $p_{s_1,t_1}$ ($p_{s_2,t_2}$).
\edefin

For any unique shortest path system $(V,P)$, it can be easily verified that its VC-dimension is $2$~\cite{Goldberghd2,Tao11}.
Thus, if a graph contains only one unique shortest path system, then, the bound described in Theorem~\ref{minsizebound} can be directly  derived (following the $\epsilon$-net theorem~\cite{Haussler:1986}).
However, in our problem, there can be many different shortest paths between any given pair of vertices.
To deal with this problem, we make the following observation:

\blemma {\bf (Existence of USPS)}
\label{uspsexist}
Given any graph $G=(V,E)$, there exists a unique shortest path system $P$ in $G$.
\elemma
\bproof
We prove this lemma by induction on the edge size of graph $G$.
We first assume that when a graph $G$ has $|E|=N$ edges, it has a unique path system $P$.
Now, we add a new edge $e=(x,y)$ in $E$ (the new edge can introduce a new vertex, the new graph is denoted as $G^\prime$),
Then, we first drop all the $p_{u,v} \in P$ ($P$ is in $G$), such that $p_{u,v}$ is not the shortest path between $u$ an $v$ any more, i.e., $|p_{u,v}| >d(u,v | G^\prime)$.
Clearly, the remaining $P$ is still a unique shortest path system.
Now, for the dropped vertex pair $u$ and $v$, we must be able to construct a new shortest path between $u$ and $v$ using edge $e=(x,y)$ as follows: $p_{u,x} \cup (e=(x,y)) \cup p_{y,v}$, where $p_{u,x}$ and $p_{y,v}$ belong to the remaining $P$.
By adding those new shortest paths to $P$, we claim $P$ is the unique shortest path system containing a shortest path between any vertex pairs in $G^\prime$.
This is because for any vertex pair $s$ and $t$, either they has a shortest path which does not contain new edge $e$ or contains. For both cases, their shortest path is uniquely defined in the new path system.
\eproof

Basically, we can always extract a USPS from a general graph even when there are more than one shortest path between any pair of vertices.  Combining Lemma~\ref{gatesetnet} and ~\ref{uspsexist}, we now can prove Theorem~\ref{minsizebound}.

\noindent{\bf Proof Sketch of Theorem~\ref{minsizebound}}:
By lemma \ref{uspsexist}, for any graph $G=(V,E)$, we have unique shortest path systems $P_{\epsilon}$ and $P^\ast_{\epsilon}$,
because they are subsets of general USPS $P$ with all possible length.
Now, for the set system $(V,P^\ast_{\epsilon})$, we know that: 1) its VC-dimension is at most 2~\cite{Goldberghd2,Tao11};
2) $\frac{\epsilon-1}{|V|}$-net on this set system is a gate-vertex set by lemma~\ref{gatesetnet}.
Using $\epsilon$-net theorem, we have a gate-vertex set (i.e., $\frac{\epsilon-1}{|V|}$-net) of size $O(\frac{|V|}{\epsilon-1} \log \frac{|V|}{\epsilon-1})$.
Moreover, the size of minimum gate-vertex set is no larger than any gate-vertex set.
Putting both together, the theorem follows.
\eproof



\begin{figure}
\centering
\begin{tabular}{c}
\includegraphics[width=0.7in, height=0.7in]{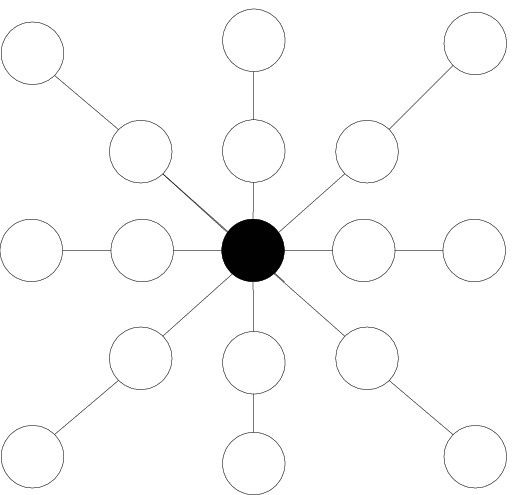}
\end {tabular}
\caption {Gate-vertex set ($\epsilon=3$)}
\label{lowerboundexamp}
\end{figure}

\noindent{\bf Lower Bound:}
\comment{
Theoretically, given input graph $G=(V,E)$ with parameter $\epsilon$, we are able to achieve the bound of gate-vertex set claimed before by following 2 steps:
1) compute an unique shortest path system (USPS) $P$ which is used for constructing the set system $(V,P^\ast_\epsilon)$;
2) randomly select $O(\frac{|V|}{\epsilon-1} \log \frac{|V|}{\epsilon-1})$ vertices to be gate-vertex set (i.e., $\frac{\epsilon-1}{|V|}$-net on $(V,P^\ast_\epsilon)$) (or we may apply the hitting set approach~\cite{Goldberghd2}).
However, it remains an open problem on how to construct a USPS efficiently. }
The lower bound of the minimum gate-vertex set can be arbitrarily small.
For example,  in Figure~\ref{lowerboundexamp}, minimum gate-vertex set is only central vertex, and no gate vertex is needed for any graph with diameter less than $\epsilon$.
In this case, even a gate-vertex set of size $O(\frac{|V|}{\epsilon-1} \log \frac{|V|}{\epsilon-1})$ is obtained,
we still cannot decide {\em how good it is compared to the minimum gate-vertex set.}

\comment{
the tightness of aforementioned upper bound is unknown
due to the non-existence of non-trivial lower bound, in contrast to what one might intuitively think (i.e., $|V|/\epsilon$).
We observe that, gate-vertex set for general graph could be arbitrarily small, even none in some extreme cases.
Considering the example
Hence,

However, several unresolved issues lead to the failure of this method in practical.
First, the method for constructing USPS mentioned in the proof of lemma~\ref{uspsexist} is very complicated and time-consuming, and even impossible to be performed in small graph.
Moreover, it is rather difficult to determine a sample of ``good'' size to be gate-vertex set,
because $O(\frac{|V|}{\epsilon-1} \log \frac{|V|}{\epsilon-1})$ is not exact integer and could be unnecessarily large.
More importantly, we cannot quantitatively measure how good the gate-vertex set generated by this method is comparing to smallest one.
In other words,
}

\comment{
The improvement of performance on many applications highly relies on the size reduction of simplified graph over original graph.
For instance, in distance computation, smaller simplified graph is, more search space will be reduced eventually leading to larger performance gain.
No performance improvement can be guaranteed if gate-vertex set of ``good'' size cannot be guaranteed.
Therefore, a gate-vertex set of ``small'' size is urgently required.
By Theorem~\ref{nphardmgslocal}, the problem of minimum gate-vertex set discovery is NP-hard and no polynomial algorithm exists unless $P=NP$.
In this case, a natural question arises: {\em can we propose a deterministic algorithm with guaranteed approximation bound for minimum gate-vertex set problem?}
We will address it in next section.
}

\comment{
Even we have upper bound $O(\frac{|V|}{\epsilon-1} \log \frac{|V|}{\epsilon-1})$ for size of minimum gate-vertex set,
since its lower bound can be arbitrarily smaller even no gate vertices are needed in some extreme case (the graph with diameter no greater than $\epsilon$).
For the shortest path query in general graphs, the search space of most of recent methods based on Dijkstra algorithm relies on the size of graph
}

\comment{

Given the input graph $G=(V,E)$, let $p_{s,t}$ be one of shortest paths between $s$ and $t$.

Let $P$ be union of all possible shortest paths where only {\bf one} shortest path for each reachable vertex pair is included ,
i.e., $P=\cup_{s,t \in V} p_{s,t}$.

To facilitate the discussion,

In addition, given parameter $\epsilon \in [0,1]$, a set $N \subseteq U$ is an $\epsilon$-net on $(U,R)$ if for any subset $X \in R$ with size no less than $\epsilon |U|$,
the set $N$ contains at least one element of $X$.
For the set system with bounded VC-dimension, the existence of small $\epsilon$-net is nicely described in $\epsilon$-net theorem by Haussler and Welzl \cite{Haussler:1986}.
To make the paper self-contained, we rephrase the theorem which will be used later as follows:

\bthm {\bf (The $\epsilon$-net theorem)~\cite{Haussler:1986}}
Let $(U,R)$ be the set system with VC-dimension at most $d$.
Then a random sample $N \subseteq U$ of size $O(\frac{d}{\epsilon} \log \frac{1}{\epsilon})$
is an $\epsilon$-net for $(U,R)$ with probability at least $1-\delta$.
\ethm

As we can see, the size of $\epsilon$-net on any set system $(U,R)$ only relies on parameter $\epsilon$ and its VC-dimension,
while is independent of the number of elements in $U$.
Roughly speaking, if the VC-dimension of one set system is bounded by some value,
its related $\epsilon$-net can be obtained with high probability by simple random sampling in theory.

To facilitate the discussion, we first define {\em Unique Shortest Path System}:

\bdefin {\bf (Unique Shortest Path System (USPS))~\cite{Goldberghd2}}
Given graph $G=(V,E)$ and a collection $Q$ of shortest paths from $G$,
we say $Q$ is unique shortest path system if for any two shortest path $p_{s_1,t_1}, p_{s_2,t_2} \in Q$,
there is no two vertices $u,v$ such that $p_{u,v}$ is subpath of $p_{s_1,t_1}$, $p^\prime_{u,v}$ is subpath of $p_{s_2,t_2}$
and $p_{u,v} \neq p^\prime_{u,v}$.
\edefin

Given this, we have the following lemma:

\blemma {\bf (VC-dimension of USPS)~\cite{Goldberghd2,Tao11}}
\label{uspsvcbound}
Given graph $G=(V,E)$ with parameter $l \ge 1$, and $P$ is unique shortest path system,
the VC-dimension of set system $(V,P)$ is at most $2$.
\elemma
\comment{
\bproof
If $\epsilon=1$, the result is trivially hold.
Otherwise, we prove it by the contradiction.
Assuming there is a shatterable vertex set $M=\{u_1,u_2,u_3\}$ of size $3$, let $u_1,...,u_2,...,u_3$ be the vertices with the same order in shortest path $p$.
In order to shatter $M$, $P$ should include one shortest path $p^\prime$ passing through $u_1$ and $u_3$ while not passing through $u_2$.
In this sense, the existence of $p$ and $p^\prime$ contradicts the definition of USPS such that there are at least two shortest paths between $u_1$ and $u_3$ (one includes $u_2$, another does not include it).
Therefore, the lemma holds.
\eproof
}

\comment{
\begin{figure}
\centering
\begin{tabular}{c}
\psfig{figure=Figures/uspsproof.eps,scale=0.4}
\end {tabular}
\caption {Example for proof of Lemma~\ref{uspsexist}}
\label{uspsproof}
\end{figure}

\begin{figure}
\centering
\begin{tabular}{c}
\psfig{figure=Figures/cases.eps,scale=0.4}
\end {tabular}
\caption {Cases in proof of Lemma~\ref{uspsexist}}
\label{cases}
\end{figure}

%

\blemma {\bf (Existence of USPS)}
\label{uspsexist}
Given graph $G=(V,E)$, there exists a unique shortest path system $P$.
\elemma
\bproof
We prove it by construction of USPS $P$ provided any $P^\prime$ including one shortest path for each vertex pair in $G$.
Let $P$ be unique shortest path system during the construction procedure.
Let $C$ be the set of vertex pair whose shortest path has been included in $P$, and $E$ denotes the set of vertex pair whose shortest path has not been included in $P$.
That is, $|C|+|E|=|V|^2$.
The basic idea of construction scheme is to maximally utilize existing shortest paths to represent each unprocessed shortest path.
We start from the empty $P$ and add one shortest path incrementally which guarantees $P$ is still USPS.
Specially, the overall scheme consists of following steps:

\noindent Step 1: remove one vertex pair $(u,v)$ from $E$, and get the corresponding shortest path $p_{u,v}$ from $P^\prime$;

\noindent Step 2: replace some segments of $p_{u,v}$ using existing shortest paths in $P$ (discuss the detail later);

\noindent Step 3: add processed $p_{u,v}$ and all its subpaths into $P$ ($C$ is updated accordingly).
Then, if $E$ is empty, USPS $P$ is generated and process terminates, otherwise, go to Step 1.

Now, we discuss the details of Step 2 to process shortest path $p_{u,v} = (u=x_0,x_1,...,x_{k}=v)$ iteratively until no segment can be replaced.
Basically, at each iteration, we scan each vertex $x$ in the shortest path from left to right and try to find anther $y$ within it such that shortest path between them exists in $P$ and its length is longest.
Let $b$ be the index of vertex $x_b$ to be processed ($b=0$ initially).
Also, we set the flag of each edge to be ``unreplaced'' at the beginning.
The procedure is outlined as follows:

\noindent 1) Considering the vertex $x_b \in p_{u,v}$, we try to find another vertex $x_e \in p_{u,v}$ such that $p_{x_b, x_e}$ has been included in $P$
and replaces largest number of edges with flag ``unreplaced''.
Replace the segment between $x_b$ and $x_e$ in $p_{u,v}$ and set those edges' flags to be ``replaced'';

\noindent 2) If $p_{x_b, x_e}$ does not exist, let $b=b+1$, otherwise, let $b=e$. Then, go to step 1) until $b \ge k$.

\noindent 3) If no replacement occurs, procedure stops, otherwise, let $b=0$ and go to step 1) to start next iteration.

Two questions we have to answer in order to ensure the correctness of construction scheme:
1) Does this procedure guarantee to stop in finite iterations?
2) Is $P$ together with processed $p_{u,v}$ and its subpaths still USPS?
For first question, replacement occurs means some ``unreplaced'' edges are replaced by existing shortest paths,
i.e, more edges are highlighted in Figure~\ref{uspsproof}.
Since the number of edges in $p_{u,v}$ is fixed, the procedure takes at most $|p_{u,v}|$ iterations to change edges' flags to ``replaced''.

For second question, assuming there are two different shortest paths for vertices $x$ and $y$
such that $p_{x,y} \subseteq p_{u,v}$ and existing shortest path $p^\prime_{x,y} \in P$.
We consider $4$ cases of $x,y$ locating at different positions of processed $p_{u,v}$
(i.e., alternate shortest path $p^\prime_{u,v} \in P$ is dashed line in Figure~\ref{cases}).
For case 3, it is easy to observe that $p^\prime_{x,y}$ instead of $p_{x,y}$ should be used to replace the segment since $|p^\prime_{x,y}| \ge |p_{x,y}|$.
For the rest of cases, they are guaranteed to be detected and resolved in the next iteration based on the procedure.
The replaced segments of processed $p_{u,v}$ (highlighted segments in Figure~\ref{uspsproof}) are essentially composed of existing shortest path by overlapping or concatenating ways.
In this way, we maintain the property of USPS of $P$ together with processed shortest path $p_{u,v}$.
\eproof

\comment{
In order to show the correctness of construction scheme, we only need to make sure that $P \cup \{p_{u,v}\}$ is still USPS.
We prove it by contradiction here.
After aforementioned process, suppose there are two different shortest paths for vertices $x$ and $y$
such that $p_{x,y} \subseteq p_{u,v}$ and existing shortest path $p^\prime_{x,y} \in P$.
We consider $4$ cases of $x,y$ locating at different positions of processed $p_{u,v}$.

\noindent Case 1: both $x$ and $y$ lie in the replaced segments (i.e., belong to existing shortest paths in $P$) (as case 1 in Figure~\ref{cases}).
According to our procedure, $p^\prime_{x,y}$ and all its subpaths are included in $P$.

Besides the shortest path $p_{x,y}=(x,x^\prime,...,y^\prime,y)$, let us assume there is alternate shortest path $p^\prime_{x,y} \in P$ (i.e., dashed line shown in Case 1 of Figure \ref{cases}).
Let $x^\prime$ and $y^\prime$ be the ending node and staring node of corresponding segment to which they belong, respectively.
According to our procedure, both $x^\prime$ and $y^\prime$ exist in the original path.
Since all $p^\prime_{x,y}$ (dashed line), $p_{x,x^\prime}$ and $p_{y^\prime,y}$ belong to $P$,
there should exist one shortest path $p^\prime_{x^\prime,y^\prime}$ belonging to $P$.
However, based on our procedure, we always can find $p^\prime_{x^\prime,y^\prime}$ staring from $x^\prime$ to replace the corresponding subpath in original path,
which makes the contradiction.
Therefore, we claim that there is no alternate shortest path exists under this condition.
}

\comment{
Assuming $p_{u,v} = (x_0,x_1,...,x_{k-1})$ is shortest path to be processed where $u=x_0$ and $v=x_{k-1}$, we set the starting point $b=0$ and
a sequence of indicators $(a_0,a_1,...,a_{k-1})$ denoting whether segments are included in $P$.
If $(x_s, ..., x_t)$ is subpath in $P$, then $a_s,...,a_t$ are set to be 1, otherwise, they are $0$ defaultly.
The procedure consists of 4 key steps:

1) We pick one vertex pair $(u,v)$ from $E$, and get the corresponding shortest path $p_{u,v}$ from $P^\prime$.

2) We try to find a shortest path $(x_b, ..., x_e) \subseteq P$ of longest length to replace subpath of $(x_b,...x_e) \subseteq p_{u,v}$ (this can be easily implemented by organizing existing $P$ in shortest path trees).
Correspondingly, indicators $a_b,...,a_{e-1}$ are set to be $1$ and starting point $b=e$.
If we cannot find any existing shortest path in $P$ starting from $x_b$, $a_b$ is set to be $0$ and $b$ is set to be $b+1$.
In this sense, the shortest path $p_{s,t}$ is represented as concatenation of segments in or not in $P$ (replaced segments are highlighted in Figure \ref{uspsproof}).

3) We add processed shortest path $p_{u,v}$ and all its subpaths whose one of endpoint's indicator is $0$. In the meanwhile, vertex pair set $C$ and $E$ are updated accordingly.

4) If $b=k-1$, we terminate the procedure, otherwise, go to step 1.

We iteratively process each shortest path by aforementioned method until uncovered vertex pair set $E$ is empty.

In order to show the correctness of our construction, we only need to confirm that $P$ with newly added shortest path $p_{u,v}$ is still USPS.
In other words, we have to guarantee that processed shortest path $p_{u,v}$ does not conflict with existing paths in $P$.
We prove it by contradiction here.
Suppose there are two different shortest paths for vertices $x$ and $y$ such that $p_{x,y} \subseteq p_{u,v}$ and existing shortest path $ p^\prime_{x,y} \in P$.
We consider $4$ cases of $x,y$ locating at different positions of processed $p_{u,v}$.

Case 1): Both $x$ and $y$ lie in the replaced segments (i.e., belong to existing shortest paths in $P$).
Besides the shortest path $p_{x,y}=(x,x^\prime,...,y^\prime,y)$, let us assume there is alternate shortest path $p^\prime_{x,y} \in P$ (i.e., dashed line shown in Case 1 of Figure \ref{cases}).
Let $x^\prime$ and $y^\prime$ be the ending node and staring node of corresponding segment to which they belong, respectively.
According to our procedure, both $x^\prime$ and $y^\prime$ exist in the original path.
Since all $p^\prime_{x,y}$ (dashed line), $p_{x,x^\prime}$ and $p_{y^\prime,y}$ belong to $P$,
there should exist one shortest path $p^\prime_{x^\prime,y^\prime}$ belonging to $P$.
However, based on our procedure, we always can find $p^\prime_{x^\prime,y^\prime}$ staring from $x^\prime$ to replace the corresponding subpath in original path,
which makes the contradiction.
Therefore, we claim that there is no alternate shortest path exists under this condition.

For the rest of 3 cases illustrated in Figure \ref{cases}, we follow the similar idea to show that no alternate shortest path exists.
Thus, utilizing our procedure, we are able to construct USPS by incrementally processing and adding the shortest path.
}
}

}

\section{Algorithms for Gate-Vertex Set Discovery}
\label{gatediscovery}


\comment{
In order to achieve that size bound, we need to precompute the unique shortest path system (USPS) which is very difficult so far.
In this section, we first introduce an fast and scalable algorithm to discover gate-vertex set. 
Since the first scheme does not does not hold any bound regarding minimum gate-vertex set,... }

Based on Theorem~\ref{minsizebound} and $\epsilon$-net theorem ~\cite{Haussler:1986}, we observe that any random sample  with size $O(\frac{|V|}{\epsilon-1} \log \frac{|V|}{\epsilon-1})$  has high probability to form a gate-vertex set but does not have a guarantee. An {\em adaptive sampling} method ~\cite{Tao11} is introduced to guarantee to find a $k$-skip cover.
The guarantee is achieved by choosing a vertex using the information gained from previously sampled vertices.
Since a $k$-skip cover can serve as a candidate for the gate-vertex set with $\epsilon=k+1$ (as stated in Lemma~\ref{kscgv}), we can utilize the adaptive sampling method to discover gate-vertex set.
However, since the lower bound of the minimum gate-vertex set can be arbitrarily small, the approximation ratio between the size of the gate-vertex set discovered by this method and the minimum one is not bounded.
In other words, this method does not necessarily produce tight gate-vertex set.
\blemma
\label{kscgv}
Given graph $G=(V,E)$,
if parameters $\epsilon$ of gate-vertex set and $k$ of $k$-skip cover satisfy condition $k = \epsilon-1$,
$k$-skip cover $V^\ast$ is a gate-vertex set.
\elemma

\bproof
We prove it by way of contradiction.
Let us assume $V^\ast$ is not gate-vertex set, meaning, there exists a vertex pair $(u,v)$ with distance $d(u,v)=\epsilon$
and we do not have one vertex $x \in V^\ast$ (note that $x \neq u,v$) such that $d(u,v)=d(u,x)+d(x,v)$.
By definition of $k$-skip cover, we guarantee to have one shortest path $p_{u,v}$ in which $V^\ast$ contains at least one vertex out of every consecutive $k$ vertices.
Therefore, for $p_{u,v}$, only starting point $u$ and ending point $v$ are allowed to be included in $V^\ast$.
However, even both $u$ and $v$ are selected,
$V^\ast$ still does not contain any vertex from subpath $p_{u^\prime, v^\prime}$ with $k$ vertices (since $p_{u,v}$ has $k+2$ vertices).
This reaches a contradiction.
\eproof

Note that a gate-vertex set with locality parameter $\epsilon=k+1$ may not be a $k$-skip cover.
Also, as we mentioned earlier, the $k$-skip cover focuses on the unique shortest path system, and since there may exist more than one shortest path between two vertices, the adaptive sampling method chooses one of such paths arbitrarily.

We propose a set-cover-based algorithm with guaranteed logarithmic bound and compare it with the adaptive sampling method.

\comment{
If $k$-skip cover can be generated, we guarantee that a gate-vertex set with locality parameter $k+1$ has been generated.
However, Lemma~\ref{kscgv} does not hold conversely, i.e., gate-vertex set is not necessarily a $k$-skip cover when $k=\epsilon-1$.
In other words, $k$-skip cover is not equivalent to gate-vertex set when $k=\epsilon-1$.
Finally, we note that the adaptive sampling method does not provide an approximation bound for our problem (the bound claimed in ~\cite{Tao11} only holds if a graph has only one unique shortest path system). }

\comment{
In general, this method is essentially heuristic-oriented and cannot offer any theoretical guarantee for the size of discovered $k$-skip cover.
Along this line, we also cannot measure how good those vertices used as gate-vertex set compared to minimum one.

In the following, we first introduce a heuristic algorithm based on $k$-skip cover~\cite{Tao11} to discover a gate-vertex set.
This algorithm is rather fast and scalable, however, no approximation bound is achieved.
Next, we propose an effective algorithm based on set cover framework to obtain a good gate-vertex set
whose size is not larger than that of minimum gate-vertex set by a logarithmic factor.
}

\comment{
\subsection{Fast Algorithm for Gate-Vertex Set}
\label{fastheu}

It is interesting to observe that $k$-skip cover is a subset of gate-vertex set with $\epsilon=k+1$.
If $k$-skip cover can be generated, we guarantee that a gate-vertex set with locality parameter $k+1$ has been generated.
An adaptive sampling method has been introduced in~\cite{Tao11} to find a $k$-skip cover, thus, we utilize it to fast discover a gate-vertex set.
Instead of random sampling, whether a vertex will be sampled depends on the information gained from previously sampled vertices.
In general, this method is essentially heuristic-oriented and cannot offer any theoretical guarantee for the size of discovered $k$-skip cover.
Along this line, we also cannot measure how good those vertices used as gate-vertex set compared to minimum one.


\comment{
We observe that gate-vertex set can be computed based on $k$-skip cover.
Under this observation, the scalable and efficient $k$-skip cover method, namely adaptive sampling \cite{Tao11}, is utilized here to generate gate-vertex set.
Note that, the discovered gate-vertex set is not guarantied to have a logarithmic bound compared to optimal one.
}

\comment{
\begin{figure}
\centering
\begin{tabular}{c}
\psfig{figure=Figures/heugatevertex.eps,scale=0.5}
\end {tabular}
\caption {Example for proof of lemma \ref{kscgv}}
\label{heugatevertex}
\end{figure}

\begin{figure}
\centering
\begin{tabular}{c}
\psfig{figure=Figures/gatevertex5.eps,scale=0.5}
\end {tabular}
\caption {Example of gate-vertex set ($\epsilon=5$)}
\label{gatevertex5}
\end{figure}
}

\blemma
\label{kscgv}
Given graph $G=(V,E)$,
if parameters $\epsilon$ of gate-vertex set and $k$ of $k$-skip cover satisfy condition $k = \epsilon-1$,
$k$-skip cover $V^\ast$ is a gate-vertex set.
\elemma
\bproof
We prove it by contradiction.
Let us assume $V^\ast$ is not gate-vertex set, meaning, there exists a vertex pair $(u,v)$ with distance $d(u,v)=\epsilon$
and we do not have one vertex $x \in V^\ast$ (note that $x \neq u,v$) such that $d(u,v)=d(u,x)+d(x,v)$.
By definition of $k$-skip cover, we guarantee to have one shortest path $p_{u,v}$ in which $V^\ast$ contains at least one vertex out of every consecutive $k$ vertices.
Therefore, for $p_{u,v}$, only starting point $u$ and ending point $v$ are allowed to be included in $V^\ast$.
However, even both $u$ and $v$ are selected,
$V^\ast$ still does not contain any vertex from subpath $p_{u^\prime, v^\prime}$ with $k$ vertices (since $p_{u,v}$ has $k+2$ vertices).
This reaches a contradiction.
\eproof

Clearly, $k$-skip cover is a subset of gate-vertex set if $k = \epsilon-1$.
However, the lemma does not hold conversely, i.e., gate-vertex set is not necessarily a $k$-skip cover when $k=\epsilon-1$.
In other words, $k$-skip cover is not equivalent to gate-vertex set when $k=\epsilon-1$.
More results on their relationship can be found in the full technical report~\cite{}.

\comment{
To gain better understanding of difference between both concepts, we consider the most simple scenario -- single path.
As shown in Figure~\ref{gatevertex5}, with $\epsilon=5$, only one gate vertex (highlighted one) is needed.
However, this is insufficient to be a $4$-skip cover yet since it does not include any vertex from first $4$ consecutive vertices.
Hence,
}


\input{Figures/FastHeu.tex}

Given this, we can apply adaptive sampling method for finding $k$-skip cover to discover gate-vertex set with $\epsilon=k+1$.
First of all, let $N_k(u)$ be a set of vertices $v$ with distance $d(u,v)=k$, namely $k$-hop neighbor set.
We denote $T_k(u)$ to be $k$-hop shortest path tree rooted at $u$ which is generated by breath-first search traversal and all $u$'s $k$-hop neighbors are its leaves.
We say $T_k(u)$ is covered by a set of vertices $V^\ast$ if every path from root $u$ to leaves within $T_k(u)$ goes through at least one vertex in $V^\ast$.
Algorithm \ref{alg:fastgate} shows the outline of this method.
Specially, let $V^\ast$ be gate-vertex set (initially $V^\ast = \emptyset$) (Line 1).
At the beginning, we sort the vertices based on the decreasing order of their corresponding degree.
Following this order, we pick one vertex $u$ and build $(\epsilon-2)$-hop shortest path tree $T_{\epsilon-2}(u)$ by performing breath-first search until the vertex with $\epsilon-1$ hops away from root is visited.
Then, we check whether $V^\ast$ covers all shortest paths from root $u$ to leaves in $T_{\epsilon-2}(u)$ (Line 4).
This can be easily achieved by another breath-first search traversal on $T_{\epsilon-2}(u)$.
If one of above shortest paths is not covered, vertex $u$ is added into $V^\ast$ (Line 6).
The procedure proceeds until all vertices are processed.
Since we apply the breadth-first search process, the overall computational time is linear with respect to the edge size together with $O(|V| \log |V|)$ cost for vertex sorting. More detailed analysis can be found in ~\cite{Tao11}.

\comment{
The computational complexity is
\noindent{\bf Computational Complexity: }
The first step of Algorithm \ref{alg:fastgate} to sort all vertices takes $O(|V| \log |V|)$ time.
For the main loop, at each iteration, it consists of two key steps:
1) perform breath-first search to build shortest path tree $T_{\epsilon-2}(u)$;
and 2) traverse $T_{\epsilon-2}(u)$ again to check whether all shortest paths with $\epsilon-1$ vertices from this tree are covered by $V^\ast$.
Since breath-first search is conducted in both steps,
the running time of both steps are essentially the same which is $O(|N_{\epsilon-2}(u)| + |E_{\epsilon-2}(u)|)$,
where $|N_{\epsilon-2}(v)|$ ($|E_{\epsilon-2}|$) is the number of vertices (edges) in the $v$'s $\epsilon-2$ neighborhood;
Clearly, the main loop processed each vertex once, thus, it takes $O(\sum_{u \in V} (|N_{\epsilon-2}(u)| + |E_{\epsilon-2}(u)|))$ in total.
Putting both together, overall time complexity is $O(|V| \log |V| + \sum_{u \in V} |E_{\epsilon-2}(u)|)$.

As locality parameter $\epsilon$ is supposed to be small integer, the size of edge set $E_{\epsilon-2}(u)$ traversed by BFS is quite small,
suggesting that this algorithm is efficient.
On the other hand, during the entire procedure, only one $(\epsilon-2)$-hop shortest path tree $T_{\epsilon-2}(u)$ is maintained in the memory when the algorithm processes vertex $u$.
We can see, the usage of memory is rather stable and small due to the size of shortest path tree with small $\epsilon$.
Even so, the aforementioned algorithm is essentially a sampling method which does no hold any approximation bound with respect to minimum gate-vertex set.
}

Finally, we note that by theoretical study in Subsection~\ref{sizegateset},
only gate-vertex set discovered by sampling vertices from unique shortest path system (USPS) has size bound of $O(\frac{|V|}{\epsilon-1} \log \frac{|V|}{\epsilon-1})$.
However, we cannot guarantee to generate USPS by BFS in Algorithm~\ref{alg:fastgate} (Line 4),
meaning that the size of discovered gate-vertex set is not necessarily bounded by $O(\frac{|V|}{\epsilon-1} \log \frac{|V|}{\epsilon-1})$.
}



\subsection{Set-Cover Based Approach}
\label{setcovergate}

\comment{
According to the complexity analysis in the previous subsection, aforementioned method is rather fast and memory-friendly which can be scaled to very large graphs.
However, the main issue is that this method does not necessarily discover a gate-vertex set of ``good'' size which is comparable to optimal solution by guaranteed factor.
On the other hand, as we mentioned, finding minimum $k$-skip cover is still open problem.
Considering the relationship between $k$-skip cover and gate-vertex set in lemma~\ref{kscgv}, the approximation algorithm for gate-vertex set potentially leads to approximation solution for $k$-skip cover.
We will demonstrate this in Section~\ref{ksipcover}.


To resolve this concern, by MGS problem definition using local-condition,}

We propose an effective algorithm based on set cover framework to discover gate-vertex set with logarithmic bound.
Specially, we transform the minimum gate-vertex set  discovery problem (MGS) to an instance of set cover~\cite{wsetcover79} problem:
Let $U=\{(u,v) | d(u,v) = \epsilon\}$ be the ground set, which includes all the non-local pairs with distance equal to $\epsilon$.
Each vertex $x$ in the graph is associated with a set of vertex pairs $C_x=\{(u,v) | d(u,v)=d(u,x)+d(x,y) =\epsilon\}$,
where $C_x$ includes all of the non-local pairs with distance equal to $\epsilon$ and there is a \textcolor{black}{shortest path} between them going through vertex $x$.
Given this, in order to discover the minimum gate-vertex set, we seek {\em a subset of vertices $V^\ast \in V$ to cover the ground set,
i.e., $U = \bigcup_{v \in V^\ast} \textcolor{black}{C_v}$, with the minimum cost $|V^\ast|$.}
Basically, $V^\ast$ serves as the index for the selected candidate sets to cover the ground set.

\bthm
The \textcolor{black}{minimum} solution $V^\ast$ for the above set-cover instance is a \textcolor{black}{minimum} gate-vertex set of graph $G$ with parameter $\epsilon$.
\ethm

Its proof can be easily followed by Definition~\ref{gatevertexsetlocal}.
The \textcolor{black}{minimum} set cover problem is NP-hard, and we can apply the classical greedy algorithm~\cite{wsetcover79} for this problem:
{\em
Let $R$ records the covered pairs in $U$ (initially, $R=\emptyset$).
For each possible candidate set $C_x=\{(u,v) | d(u,v)=d(u,x)+d(x,y) =\epsilon\}$ discussed above, we define the price of $C_x$ as:
\[\gamma(C_x)=\frac{1}{|C_x \setminus R|}\]
At each iteration, the greedy algorithm picks the candidate set $C_x$
with the minimum $\gamma (C_x)$ (the cheapest price) and put its corresponding vertex $x$ in $V^\ast$.
Then, the algorithm will update $R$ accordingly, $R=R\cup C_x$.
The process continues until $R$ completely covers the ground set $U$ ($R=U$), which contains all non-local pairs with distance equal to $\epsilon$.
It has been proved that the approximation ratio of this algorithm is
$ln(|U|)+1$~\cite{wsetcover79}.}

\noindent {\bf Fast Transformation: }
In order to adopt the aforementioned set-cover based algorithm to discover the gate-vertex set, we first have to generate the ground set $U$ and each candidate subset $C_x$ associating with vertex $x$.
Though we only need the non-local pairs with distance $\epsilon$, whose number is much smaller than all non-local pairs,
the straightforward approach still needs to precompute the distances of each pair of vertices with distance no greater than $\epsilon$,
and then apply such information to generate each candidate set.
For large unweighted graphs, such computational and memory cost can still be rather high.

Here, we introduce an efficient procedure which performs a local BFS for each vertex to visit only its $\epsilon$-neighborhood and during this process to collect all information needed for constructing the set-cover instance ($U$ and $C_x$, for $x\in V$).
Specifically, for each local BFS starting from a vertex $u$, it has the following two tasks: 1) it needs to find all the vertices which is exactly $\epsilon$ distance away from $u$, and then add them into $U$;
and 2) for each such pair $(u,v)$ ($d(u,v)=\epsilon$), it needs to identify all the vertices $x$ which can appear in a shortest path from $u$ to $v$.
In order to achieve these two tasks, we again utilize the basic recursive property of the shortest-path distance:
{\em Let vertex $y$ to $u$'s distance be $d$ and  $z \neq u$ be vertex whose distance to $u$ is $d-1$,
we know all the intermediate vertices appearing in at least one shortest path from $u$ to $z$ (denoted as $I(z)$).
Then, all the intermediate vertices on \textcolor{black}{shortest paths} from $u$ to $y$ can be written as $\bigcup_{(z,y)\in E} (I(z) \cup \{z\})$. }
Based on this property, we can easily maintain $I(x)$ for each vertex $x$ such that $1<d(u,x) < \epsilon$; when $d(u,x)=1$, $I(x) =\emptyset$.
Since BFS visits $u$'s $\epsilon$-neighborhood in a level-wise fashion, when it reaches the $\epsilon$ level, where each vertex $v$ is $\epsilon$ distance to $u$, we not only get each targeted pair $(u,v)$, but also get $I(v)$, which we can easily use for producing the candidate set:
for each $x \in I(v)$, we add $(u,v)$ to $C_x$.

Algorithm \ref{alg:bfssetcover} sketches the BFS-based algorithm for constructing the set cover instance.
Especially, set $I(v)$ is computed in Line $5$, and when BFS reaches the $\epsilon$ level (Line $11$), it adds $(u,v)$ to the ground set $U$ (Line $12$) and to each $C_x$ ($x \in I(v)$) (Line $13$).
The algorithm will be invoked for each vertex $u$ in the graph.
Finally, Figure~\ref{bfsexample} illustrates a simple running example of Algorithm~\ref{alg:bfssetcover} for vertex $u$ with $\epsilon=3$.

\begin{algorithm}
{\small
\caption{BFSSetCoverConstruction($G=(V,E)$,$\epsilon$,$U$,$u$)}
\label{alg:bfssetcover}
\begin{algorithmic}[1]
\STATE $I(u) \leftarrow \emptyset$; $level(u) \leftarrow 0$; $Q \leftarrow \{x\}$ \COMMENT{queue for BFS};
\WHILE{$Q \neq \emptyset$}
    \STATE $u \leftarrow Q.pop()$;
    \IF {$level(v) \geq 2$ \COMMENT{$d(u,v)\geq 2$}}
         \STATE $I(v) \leftarrow \bigcup_{(x,v) \in E \wedge level(x)+1=level(v)} I(x) \cup \{x\}$
    \ENDIF
    \FORALL{$v \in Neighbor(u)$ \COMMENT{$(u,x) \in E$}}
        \IF{$v$ is not visited}
            \IF{$level(v) \textcolor{black}{<} \epsilon$ \COMMENT{$d(u,v) \textcolor{black}{<} \epsilon$}}
                \STATE $Q.push\_back(v)$;
            \ELSE
                \STATE $U \leftarrow U \cup \{(u,v)\}$;
                \STATE $\forall x \in I(v)$, $C_x \leftarrow C_x \cup \{(u,v)\}$;
            \ENDIF
        \ENDIF
    \ENDFOR
\ENDWHILE
\end{algorithmic}
}
\end{algorithm}

\begin{figure}
    \centering
    \mbox{
        \subfigure[Example for Algorithm \ref{alg:bfssetcover}] {\includegraphics[scale=0.5]{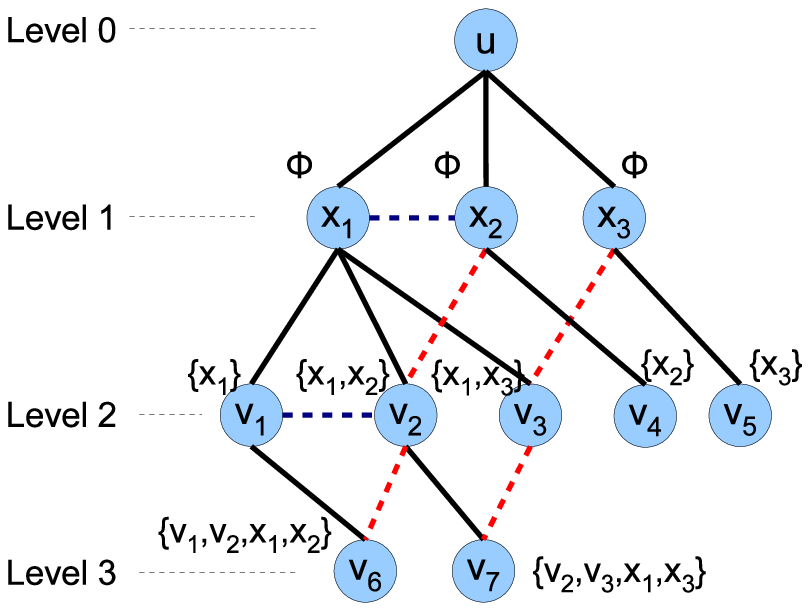} \label{fig:bfs}}
        \subfigure[Example for Set Cover]{\includegraphics[scale=0.5]{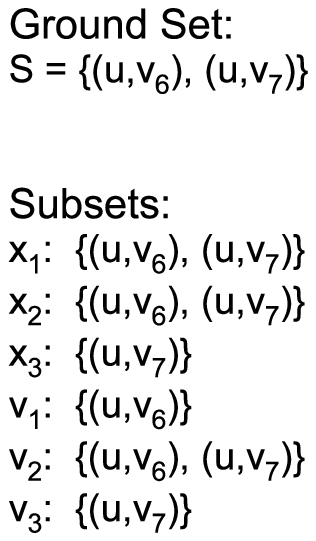}\label{fig:setcover}}
    }

    \caption{Example for Gate Discovery Algorithm ($\epsilon=3$)}
    \label{bfsexample}
\end{figure}

\comment{
\begin{figure}
\centering
\begin{tabular}{c}
\psfig{figure=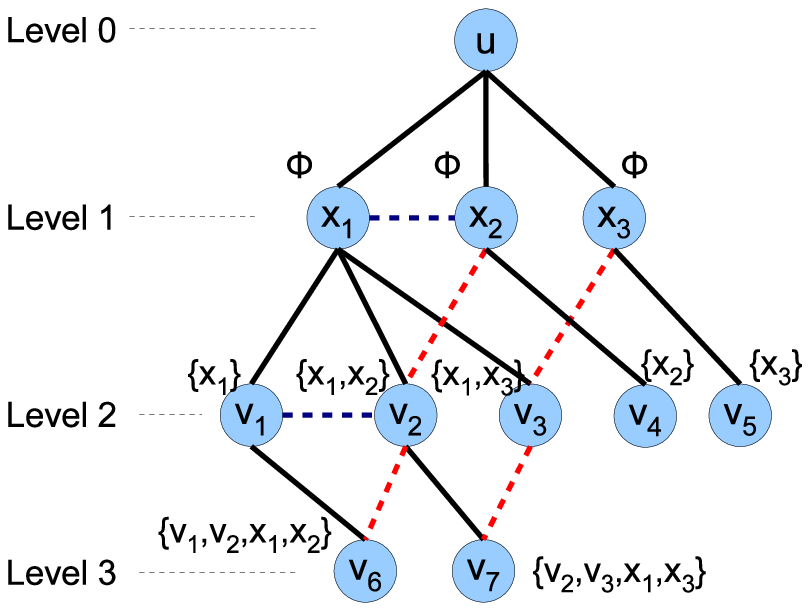,scale=0.6}
\end {tabular}
\caption {Illustrative Example for Algorithm \ref{alg:bfssetcover}}
\label{bfsexample}
\end{figure}
}

\noindent{\bf Computational Complexity:}
The overall set-cover based mining algorithm for discovering gate-vertex set includes two key steps:
1) Constructing set-cover instance (Algorithm~\ref{alg:bfssetcover}) and 2) the greedy set-cover discovering algorithm.
The first step for collecting ground set and each candidate set takes $O(\sum_{v \in V} (|N_{\epsilon}(v)|^2+|E_{\epsilon}(v)|))$, where $|N_{\epsilon}(v)|$ ($|E_{\epsilon}(v)|$) is the number of vertices (edges) in the $v$'s $\epsilon$-neighborhood.
For the greedy set cover procedure, by utilizing the speedup queue technique~\cite{Schenkel04,DBLP:conf/sigmod/JinXRF09},
we only need to visit $d \ll |V|$ vertices in the queue (i.e., all vertices are ranked in ascending order in the queue),
and each step takes $O(d(\log |V| + 1))$ time to exact and update the queue.
As greedy procedure has $O(|V^\ast|)$ steps, it takes $O(d |V^\ast| (\log |V| +1))$ in total.
Putting together, the overall algorithm's time complexity is $O(d |V^\ast| \log |V|  + \sum_{v \in V} |N_{\epsilon}(v)|^2)$.

\comment{
\noindent{\bf Memory-Saving Technique:}
A potential issue of our gate discovery algorithm is that it requires materialization of the vertex pairs whose distance is no greater than $\epsilon+1$ (i.e., ground set),
which is needed for greedy set cover procedure.
If graph diameter is very close to $\epsilon$, the size of vertex pairs can be very close to the size of transitive closure, which is hard to be hold in the main memory.
To handle large graphs, we adopt partial set cover strategy.
Instead of trying to cover all the shortest pairs with distance $\epsilon+1$, we first pick a set of vertices $V^\ast_b$ with high betweenness as our gates.
The vertex betweenness based on the shortest path no longer than $\epsilon$ can be easily estimated during BFS traversal.
This actually follows the basic idea of greedy set cover procedure, because the vertex with high betweenness tends to cover more shortest pairs in the greedy procedure.
Then, only the vertex pairs which have not covered by $V^\ast_b$ are considered in the set cover framework.
Supposing a large portion of vertex pairs are covered by the vertex with high betweenness, the size of ground set in set cover framework would be significantly reduced,
which eventually avoids the memory bottleneck.
}

\comment{

\bthm({\bf NP-hardness of Minimal Gate-Vertex Set Problem})
The minimal gate-vertex set problem is NP-hard.
\ethm
\bproof
We can reduce the classical vertex cover problem (NP-hard) to this problem.

\eproof

\bproof
We can reduce the classical set cover problem (NP-hard) to this problem.
In the set cover problem, we have a ground set $U=\{u_1, u_2, \cdots, u_n\}$ which needs to be covered; and we have a list of candidate set $C_1, C_2, \cdots C_K$, which are subsets of $U$. We seek the minimal number of candidate sets to cover $U$, i.e.,$C_{i1} \cup C_{i2} \cup \cdots \cup C_{ik}=U$.
Now, for the set cover instance, we build a ($4|U|+1$)-partite graph $G=(X_1 \cup X_2 \cup \cdots X_{2|U|}\cup C \cup Y_{2|U|} \cup \cdots \cup Y_2 \cup Y_1, E)$, where $X_i=Y_i=U$ ($1\leq i \leq 2|U|$) (each element in the ground set is uniquely copied exactly once into all $X_i$ and $Y_i$ ($1 \leq  I \leq 2|U|$) and each candidate set corresponding to a vertex in $C$.
The edge set $E$ is the union of $4|U|$ edge sets: $E=\bigcup_{1 \leq j \leq 2|U|-1} E_j \cup E_{xc} \cup E_{cy} \cup \bigcup_{1 \leq j \leq 2|U|-1} E_j^\prime$, where $E_j \subseteq X_j \times X_{j+1}$ ($1 \leq j \leq 2|U|-1$), $E_{xc} \subseteq X_{2|U|} \times C$, $E_{cy} \subseteq C \times Y_{2|U|}$, and $E_j^\prime \subseteq Y_{j} \times Y_{j+1}$ ($1 \leq j \leq 2|U|-1$).
Specifically, edge sets $E_j$ and $E_j^\prime$ ($1\leq j \leq 2|U|-1$)  are similarly defined and all have only $|U|$ edges, which link each vertex in $X_j$ ($Y_j$) with its corresponding ``twin'' vertex in $X_{j+1}$($Y_{j+1}$);
edge set $E_{xc}$ and $E_{cy}$ are similarly defined: if an element $u_i \in C_j$, then there is an edge in $E_{xc}$ ($E_{cy}$) linking $u_i$'s corresponding vertex in $X_{2|U|}$ ($Y_{2|U|}$)  to $C_j$'s corresponding vertex in $C$.
Figure~\ref{fig:npfigure} illustrates the construction from a set cover problem instance.
Clearly, this construction only takes polynomial time with respect the size of the set cover problem.

Given this, we claim that the minimal gate-vertex set with parameter $\epsilon=2|U|$ corresponds the solution for  the minimal set cover problem.
To observe this, we need the following lemmas:

\blemma
The distance between any two vertices in $C$ is less than $2|U|$; for any pair with their distance is higher than $2|U|$, their shortest path must go through at least one vertex in $C$.
\elemma

\blemma
The minimal gate-vertex set contains only vertices in $C$.
\elemma
\bproof
We prove it by way of contradiction.
Suppose $V^\ast$ is minimal and $V^\ast=X^\ast \cup C^\ast \cup Y^\ast$, where $X^\ast \subseteq \bigcup X_j$, $Y^\ast \subseteq \bigcup Y_j$, and $C^\ast \subseteq C$, and $X^\ast \neq \emptyset$ or $Y^\ast \neq \emptyset$.
Without loss of generality, let $|X^\ast| \leq |Y^\ast|$ and let $Y(X^\ast)$ to the corresponding symmetric vertices in $\big Y_i$.
Note that the constructed graph is symmetric with respect to $C$.
Due to the symmetric property, we can see $V^\prime=X^\ast \cup C^\ast \cup Y(X^\ast)$ must also be a gate-vertex set and $|V^\prime$ \leq |V^\ast|$.
Let $x \in X^\ast$ and $Y(x) \in Y^\ast$ ($Y(x)$ is $x$'s symmetric vertex in $Y$).
Let $c \in C$ be
\eproof
\eproof
\begin{figure}[t]
\includegraphics[width=3.3in]{Figures/NP.eps}
  \caption{\small Constuction for Reducing Set-Cover to Minimal Gate-Vertex Set }
  \label{fig:npfigure}
\end{figure}

}

\comment{
\bdefin{\bf (Minimal Edges Discovery Problem)}
\label{edgediscovery}
Given an undirected graph $G=(V,E)$, the parameter $\epsilon$ and its corresponding gates $V^\ast$,
we want to find a set of weighted edges $E^\ast$ building the connections among $V^\ast$, and preserving the property:
\beqnarr
 d_G(u,v)= \nonumber \\
  min_{d_G(u,x) \le \epsilon \wedge d_G(y,v) \le \epsilon \wedge x,y \in V^\ast} (d_G(u,x) + d_{G^\ast}(x,y) + d_G(y,v)) \nonumber
\eeqnarr
where the number of edges $|E^\ast|$ is minimal.
\edefin

}

\comment{
The simplified graph is defined by using the vertex set $V^\ast$ and the edge set $E^\ast$.
To provide more flexibility, we also generalize our problem with exact distance preserving to the graph simplification problem with approximate distances.
Intuitively, more approximation error is allowed smaller the size of graph is.
By selecting suitable approximation ratio, users can adjust the trade-off between the simplicity of resulting graph and the accuracy of shortest distances.

\bdefin{\bf (Graph Simplification Problem with Approximate Distance)}
\label{agsprobdef}
Given an undirected graph $G=(V,E)$, the parameter $\epsilon$ and $\delta$,
its simplified graph $G^\ast=(V^\ast, E^\ast, W)$ ($V^\ast \subseteq V$)is an undirected weighted graph maintains following properties:

\noindent 1) $d_{G^\ast}(u,v) \le (1+\delta) d_G(u,v)$;

\noindent 2) $min_{d_G(u,x) \le \epsilon \wedge d_G(y,v) \le \epsilon} (d_G(u,x) + d_{G^\ast}(x,y) + d_G(y,v)) \le (1+\delta) d_G(u,v)$;

where both the number of vertices $|V^\ast|$ and the number of edges $|E^\ast|$ are minimal.
\edefin

\bdefin{\bf (Gates Discovery Problem)}
\label{agatediscovery}
Given an undirected graph $G=(V,E)$ and the parameter $\epsilon$, its corresponding gates are a subset of vertices $V^\ast \subseteq V$ keeping the following property:
\beqnarr
  min_{d_G(u,x) \le \epsilon \wedge d_G(y,v) \le \epsilon \wedge x,y \in V^\ast} (d_G(u,x) + d_{G^\ast}(x,y) + d_G(y,v)) \nonumber \\
  \le (1+\delta) d_G(u,v) \nonumber
\eeqnarr
where the number of gates $|V^\ast|$ is minimal.
\edefin

\bdefin{\bf (Minimal Edges Discovery Problem)}
\label{aedgediscovery}
Given an undirected graph $G=(V,E)$, the parameter $\epsilon$ and its corresponding gates $V^\ast$,
we want to find a set of weighted edges $E^\ast$ building the connections among $V^\ast$, and preserving the property:
\beqnarr
  min_{d_G(u,x) \le \epsilon \wedge d_G(y,v) \le \epsilon \wedge x,y \in V^\ast} (d_G(u,x) + d_{G^\ast}(x,y) + d_G(y,v)) \nonumber \\
  \le (1+\delta) d_G(u,v) \nonumber
\eeqnarr
where the number of edges $|E^\ast|$ is minimal.
\edefin
}

\comment{
\bobserv
Let $u$, $v$, $x$ and $y$ be the vertices of graph $G$, and $x$ and $y$ are distinct from $u$ and $v$.
If there exits a shortest path $SP_{u,v}=(u\codts x \cdots y \cdots v)$ contains $x$ and $y$, then the subpath $P_{x,y}=(x \cdots y) \subset SP_{u,v}$ is a shortest path.
\eobserv

This inspires to
Recall the procedure of computing shortest distance based on $\epsilon-neighbors$ and simplified graph,
we have to find out the appropriate gates $x$ and $y$ which are contained in the source and destination's $\epsilon-neighbors$, respectively.
How to find the smallest number of such vertices is also the goal of {\em Gates Discovery Problem}.
Interestingly, for each vertex $u$ in the graph, we observe that if we can identify a subset of vertices from $\epsilon-neighbors$ as its {\em gates} which can lead to the vertices $v^\prime$ where their distances $d_G(u,v^\prime)$ is $\epsilon+1$,
we can restore the distance from $u$ to any its reachable vertex $v$.

Before we formally summarize the above property, we first introduce the definition of {\em coverage} in terms of shortest distances:

\comment{
\bdefin
Let $u$, $v$ and $x$ be the vertices of graph $G$.
If $x$ lies in one of the shortest path $SP_{u,v}$ between $u$ and $v$ (i.e., $d_G(u,v) = d_G(u,x)+d_G(x,v)$),
then we say $x$ covers the shortest distance between $u$ and $v$.
\edefin
}
}

\comment{
We focus on the shortest distance no less than $\epsilon+1$.
For any shortest path $P=(v_1,...,v_k)$ with the length $k > \epsilon$, it contains the shortest path $SP_1=(v_1,...,v_{\epsilon})$ and $SP_2=(v_{k-\epsilon+1}, ..., v_k)$.
Since we have guaranteed that one vertex $x_0 \in L(v_1)$ lies in the path $SP_1$ and one vertex $y_0 \in L(v_k)$ belongs to the path $SP_2$,
we have two vertices $x_0$ and $y_0$ in the shortest path $P$.
Now, we consider the shortest path between $x_0$ and $y_0$.
Its shortest distance has two cases: 1) $d_G(x_0,y_0) \le \epsilon$; or 2) $d_G(x_0,y_0) > \epsilon+1$.

\noindent{Case 1:} $x_0$ and $y_0$ are each other's $\epsilon-neighbors$, thus the shortest distance can be calculated directly.
Then, we compute the shortest distance by concatenating three shortest distances as follows:
\[ d_G(v_1,v_k) = d_G(v_1, x_0) + d_{G}(x_0,y_0) + d_G(y_0,v_k) \]

\noindent{Case 2:} following similar idea, we always have one vertex $x_1 \in L(x_0)$ and one vertex $y_1 \in L(y_0)$ belonging to shortest path $SP_{x_0,y_0}$.
In this way, we convert the problem of computing shortest distance $x_0$ and $y_0$ to computing the distance between $x_1$ and $y_1$.
Since both $x_1$ and $y_1$ lie in the shortest path $SP_{x_0,y_0}$, we make sure that $x_1$ and $y_1$ also belong to shortest path $SP_{u,v}$ and always have $d_G(x_0,y_0) = d_G(x_0,x_1) + d(y_1, y_0)$.
We recursively find $x_{i+1} \in L(x_i)$ and $y_{i+1} \in L(y_i)$ that lie in the shortest path $SP_{u,v}$ (case 2) until the distance $d_G(x_m,y_m) \le \epsilon$ (case 1).
That is, the distance between $x_i$ and $y_i$ can be formulated as a recursive expression:
$d_G(x_i,y_i) = d_G(x_i,x_{i+1}) + d_G(x_{i+1},y_{i+1}), d_G(y_{i+1}, y_i)$ if $d_G(x_{i+1},y_{i+1})\ge \epsilon$.
Given this, we concatenate all shortest distances through $x_i$ and $y_i$ to compute $d_G(u,v)$ as follows:
\beqnarr
 d_G(v_1,v_k) = d_G(v_1, x_0) + \sum_{i=1}^m d_G(x_{i-1},x_i) +  d_{G}(x_m,y_m) \nonumber \\
        + \sum_{i=1}^m d_G(y_{i-1},y_i) + d_G(y_0,v_k) \nonumber
\eeqnarr
where $d_G(x_m,y_m)$ has been demonstrated in case 1.}

\comment{each vertex $u$, there is a subset of vertices $L(u)$ such that for any vertex $x \in L(u)$, $0<d(u,x) \leq \epsilon$, and for any vertex $v$ which is $\epsilon+1$ distance to $u$ ($d(u,v) = \epsilon+1$), there is a vertex $x \in L(u)$, such that $d(u,v)=d(u,x)+d(x,v)$.
We would like to discover these subsets of vertices, such that $|\bigcup_{u \in V} L(u)|$ is minimal ($\bigcup_{u \in V} L(u)$ is a gate-vertex set).}

\section{Algorithm for Gate Graph Discovery}
\label{gatestructure}

In this section, we study the {\em gate graph discovery} problem (Definition~\ref{gsprobdef} in
Subsection~\ref{problem}). Basically, after a gate-vertex set $V^\ast$ is discovered from graph $G$, we ask {\em how to
minimally connecting \textcolor{black}{those} gate vertices while still preserving the ability of representing non-local distances through
consecutive local pairs?} Specifically, the gate graph $G^\ast=(V^\ast,E^\ast,W)$ is a weighted graph, which guarantees
for any non-local pair $u$ and $v$ in $G$ ($d(u,v) \ge \epsilon$), $d(u,v)=$
{\small
\beqnarr min_{d(u,x) < \epsilon \wedge
d(y,v) < \epsilon \wedge x,y \in V^\ast}
              d(u,x)+d(x,y|G^\ast)+d(y,v); \nonumber
\eeqnarr}
Here $d(x,y|G^\ast)$ is the distance between $x$ and $y$ in the gate graph. To find the overall sparsest gate
graph $G^\ast$ seems to be a hard problem. Here, we develop a two-stage algorithm to try to maximally prune
non-essential edges between gate vertices.

\noindent{\bf Stage 1: Constructing Local-Gate Graph.} In the first stage, for each gate vertex $u \in V^\ast$, we
construct a {\em local-gate graph} $G^\prime$ by connecting two gate vertices only if their distance is less than
$\epsilon$: $G^\prime=(V^\ast,E^\prime,W)$, where $E^\prime=\{(u,v)| d(u,v) \textcolor{black}{<} \epsilon\} \subseteq
V^\ast \times V^\ast$, and $w(u,v)=d(u,v)$ for $(u,v) \in E^\prime$. In the next stage, we will try to sparsify the
local-gate graph by removing those non-essential edges, i.e., those edges whose removal will not affect any
\textcolor{black}{shortest-path} distance in the gate graph. Why we need only local-pairs edges in the gate graph?
Lemma~\ref{localenough} answers this question.

\blemma \label{localenough} The local-gate graph $G^\prime$ can guarantee that for any non-local pair $u$ and $v$ in
$G$ ($d(u,v) \ge \epsilon$), $d(u,v) =$
{\small
\beqnarr min_{d(u,x) < \epsilon \wedge d(y,v) < \epsilon \wedge x,y \in V^\ast}
              d(u,x)+d(x,y|G^\prime)+d(y,v); \nonumber
\eeqnarr}
\elemma

Lemma~\ref{localenough} can be derived directly from the definition of gate-vertex set. Its proof is omitted for
simplicity.

\comment{Basically, the distance of any
non-local distance can be recovered through consecutive local-pairs. The local-gate graph essentially links all the
local-pairs and thus can guarantee the non-local distances and local-distances between gate vertices can be recovered.
However, not necessarily every edge in the local-gate graph is required. For instance,  Figure~\ref{fig:cgg} shows a
subgraph centered with vertex $u$ of the local-gate graph ($\epsilon=6$). Clearly, the edges $(u,x_2)$, $(u,x_3)$, and
$(u,x_4)$ are all non-essential and can be dropped (Figure~\ref{fig:simpcgg}).

\begin{figure}
    \centering
    \mbox{
        \subfigure[$u$'s local connection]{\epsfig{figure=Figures/ggexample1.eps,scale=0.7} \label{fig:cgg}}
        \subfigure[Sparsified local-gate graph]{\epsfig{figure=Figures/ggexample2.eps,scale=0.7}\label{fig:simpcgg}}
    }
    \label{fig:cggredundancy}
    \caption{Example for Local-Gate Graph Sparsification}
\end{figure}
}

\noindent{\bf Stage 2: Edge Sparsification for Local-Gate Graph.} In this stage, for each edge in the local-gate graph,
we will determine whether removing it will change the distance between any local pair (if local pair is unchanged, so
does non-local pair based on the definition of gate vertices). This can be equivalently described in the following
condition: {\em for any edge $(u,v)$ in the local-gate graph $G^\prime$, if there is a vertex $x$ ($x \neq u, x \neq
v$), and $d(u,x)+d(x,v)=d(u,v)$, then, edge $(u,v)$ is non-essential and can be safely removed from $G^\prime$. } How
do we test this condition? Using the local-gate graph, this becomes very simple!

\blemma \label{remove} Given local-gate graph $G^\prime$, let $N(u)$ be the adjacent gate vertices of vertex $u$ in
$G^\prime$. For any edge $(u,v)$ in $G^\prime$, if there is a common vertex $x \in N(u) \cap N(v)$, such that
$w(u,x)+w(x,v)=d(u,v)$, then, removing edge $(u,v)$ from $G^\prime$ will not affect the distance between any two
vertices in $G^\prime$; if not, then, edge $(u,v)$ is essential and removing it increases the distance between at least
one pair of vertices ($u$ and $v$). \elemma

This lemma essentially utilizes the property that in the local-gate graph, any pair with distance less than
$\epsilon$ is linked through an edge in $G^\prime$ and thus we do not need to consider the situation where an edge can
be replaced by a shortest path. Here, if an edge can be replaced, there must be a shortest path with only two edges.
Given this, we can see that the pruning algorithm needs to scan the edge set of local-gate graph twice: {\em 1) it
applies Lemma~\ref{remove} to determine whether an edge can be removed and flag them; and 2) it removes all the edges
being flagged to be non-essential. } Note that we should not drop an edge immediately after we found it to be
non-essential since it can be used by testing other edges. Finally, the computational complexity of the overall edge
sparsification algorithm is $O(\sum_{v \in V} (|N_{\epsilon-1}(v)|+|E_{\epsilon-1}(v)|)+|E^\prime|)$ considering the
cost of computing the distance between local pairs of the gate vertices.

\comment{

The basic idea of our approach is: staring from one vertex $u$, we try to find a shortest path with most edges to reach its local neighbors with distance no greater than $\epsilon$.
This can be achieved by performing Dijkstra algorithm while recording the number of edges which has been passed through by this shortest path.
The sketch of algorithm is outlined in Algorithm \ref{alg:cggsparse}.
Our algorithm bears the similarity with Dijkstra algorithm, and the most important difference is the usage of arrays $previous$ and $hop$.
The variables $previous(v)$ and $hop(v)$ are used to record the previous vertex in the shortest path from source to $v$, and the maximal number of edges in the currently found shortest paths from source to $v$, respectively.
During the main loop for updating shortest distance $dist(v)$ (Line 9 to Line 26), we only check the vertices with distance no greater than $\epsilon$ (Line 11 to Line 13).
In addition, in the distance updating step (Line 15 to Line 24), $previous(v)$ and $hop(v)$ are update accordingly based on 2 cases:
1) shortest distance is updated (Line 15 to Line 18); 2) a path with the same distance is found.
In the second case, we have to compare the number edges in both paths and choose the one with more edges as its subpath to connect source.
Finally, we guarantee to have the shortest path with maximal edges connecting to source.
Then, we remove all redundant edges not included in such shortest paths from complete gate graph.

\input{Figures/GateGraphDiscovery.tex}}

\comment{
\section{Gate Graph Sparsification}
\label{sparsification}

Given aforementioned gates, a weighted undirected graph, referred to as {\em gate graph},  can be easily formed by connecting each pair of gates that are each other's $\epsilon-neighbors$.
That is, each gate connects to other gates which are its $\epsilon-neighbors$ in the graph.
Here, the weight of each edge in the gate graph represents the shortest distance between two endpoints in the original graph.
Intuitively, the gate graph could be very dense if the discovered gates are not far from each other.
As we know, the dense graph is difficult to provide a clear and understandable abstraction of original graph in many applications, such as graph visualization.
To tackle this problem, we propose a sparsification approach based on Set-Cover-with-Pairs (SCP) problem by relaxing the exact distance between each pair of gates with certain approximation error in this section.
In this sense, we provide a flexible mechanism to adjust the tradeoff between the simplicity of resulting graph and the accuracy of shortest distance computation.

To facilitate the remaining discussion, we first introduce the concept of {\em Gate Graph} and {\em $k$-hop neighbors} as follows:

\bdefin{\bf (Gate Graph)}
\label{gategraph}
Let $G=(V,E)$ be original graph, its discovered gates $V^\ast$ and parameter $\epsilon$.
The weighted graph $G^\prime=(V^\ast, E^\prime,W)$ is complete gate graph if and only if there is a edge between each pair of gates $u$ and $v$
where $d_G(u,v) \le \epsilon$ and $w(u,v)=d_G(u,v)$.
\edefin

\bdefin{\bf ($k$-hop Neighbors)}
\label{hopneighbor}
Let $G=(V,E)$ be original graph and $G^\prime=(V^\ast,E^\prime,W)$ be its corresponding gate graph with parameter $\epsilon$.
We say gate $v$ is gate $u$'s $k$-hop neighbor if and only if there is a shortest path between $u$ and $v$ with $k$ edges in $G^\prime$.
\edefin

Taking the gate $u$ as regular vertex in the original graph, it is able to directly access all its {\em gates} based on the definition of gate graph.
In other words, $u$'s immediate neighbors in the gate graph are able to cover all vertex pairs $\{(u,v_i)\}$ where $d_G(u,v_i)=\epsilon+1$.
According to lemma \ref{restorelemma}, the shortest distance between each pair of gates can be restored only utilizing gate graph.
That means, the shortest distance in the gate graph is the same with the one from original graph.
However, even the gate graph holds shortest distance restoration property, it may not be informative and clear enough to be adopted in the real-world applications due to high density.
The natural problem arises: {\em how can we simplify the gate graph by removing some edges while preserving the distance restoration property?}
Essentially, we are trying to find a minimal number of edges in the gate graph to retain gates' pair-wise shortest distances.

\bdefin{\bf (Gate Graph Simplification Problem)}
\label{ggsimpprob}
Given an undirected graph $G=(V,E)$ and its corresponding gate graph $G^\prime=(V^\ast, E^\prime, W)$,
we aim at extracting a subgraph $G^\ast=(V^\ast, E^\ast, W^\ast)$ where $E^\ast \subset E^\prime$, $W^\ast \subset W$, and satisfies the following conditions:

\noindent 1) $d_{G^\ast}(u,v) = d_{G^\prime} G(u,v)$;

\noindent 2) the number of edges $E^\ast$ is minimized.

\edefin

\begin{figure}
\centering
\begin{tabular}{c}
\psfig{figure=Figures/edgediscovery.eps,scale=0.6}
\end {tabular}
\caption {Example for Edge Selection in Gate Graph}
\label{fig:ggsparse}
\end{figure}

Following the similar idea in the gate discovery, in order to hold distance restoration property in the gate graph,
each gate $u$ should connect to some of its immediate neighbors (i.e., 1-hop neighbors) which cover all shortest distances from $u$ to its 2-hop neighbors.
Here, we use 1-hop neighbors since gate graph is a weighted graph where each edge weight corresponds to the shortest distance between two connected vertices.
Also, for each vertex $u$, its immediate neighbors are actually its $\epsilon-neighbros$ since $w(u,v) \le \epsilon$ and $v \in \epsilon-neighbor(u)$,
and the distance to its 2-hop neighbors are greater than $\epsilon+1$.
Take figure \ref{fig:ggsparse} as example.
For vertex $u$, its immediate neighbors are $\{x_1,x_2,x_3,x_4\}$ and 2-hop neighbors are $\{v_1,v_2,...,v_8\}$.
To hold the property of distance restoration, we can select edge set $\{(u,x_1),(u,x_2),(u,x_3)\}$ or $\{(u,x_1),(u,x_3),(u,x_4)\}$ to cover all 2-hop neighbors.

However, we cannot select the edge set for each vertex individually.
The central challenge here is original graph is undirected graph and each edge serves as the intermediate edge within different shortest paths from both directions.
More importantly, the effectiveness of the edge relies on its incident edges, meaning one edge should incorporate with other edges to form the shortest path.
For instance, in order to use edge $(u,x_1)$ as the edge in the shortest path from $u$ to $v_1$, we have to guarantee that the edge $(x_1,v_1)$ is also chosen.
Therefore, we cannot directly apply traditional set cover approach to find the minimal number of edges from gate graph to keep the shortest distance restoration property.

\subsection{Transformation to Set-Cover-with-Pairs Problem}
\label{transform}

In this subsection, we transform our problem to a new variant of set cover problem, referred to as {\em Set-Cover-with-Pairs} (SCP) problem, which is defined in the following.

\bdefin{\bf (Set-Cover-with-Pairs Problem ~\cite{HS05})}
\label{scpdef}
Let $U$ be the ground set and let $S = \{1, \ldots, M\}$ be a set of objects, where each object $i\in S$ has a non-negative cost $w_i$.
For every $\{i,j\}\subseteq S$, let $\mathcal{C}(i, j)$ be the collection of elements in $U$ covered by the pair $\{i, j\}$.
The objective of the set cover with pairs (SCP) problem is to find a subset $S^\prime \subseteq S$ such that
 $\mathcal{C}(S^\prime) =\bigcup_{\{i,j\}\subseteq S^\prime} \mathcal{C}(i, j) = U$ with a minimum covering cost $\sum_{i\in S} w_i$.
We refer to the special case in which each object has a unit weight ($w_i=1$) as the \textbf{cardinality SCP} problem.
\edefin

Let us first define the concept of {\em edge coverage} to be used in the SCP formulation as follows:

\bdefin{\bf (Edge Coverage)}
Let $G=(V,E)$ be original graph and $G^\prime=(V^\ast,E^\prime,W)$ be its corresponding gate graph.
There are two edges $e_1=(u,x) \in E^\prime$, $e_2=(y,v) \in E^\prime$, and vertex pair $(u,v)$ where $u$ and $v$ are each other's 2-hop neighbor.
We say $e_1$ and $e_2$ cover vertex pair $(u,v)$ if and only $x$ and $y$ are the same vertex and $d_G(u,v)=w(u,x)+w(y,v)$.
\edefin

That is, edges $e_1=(u,x)$ and $e_2=(y,v)$ constitutes the shortest path from vertex $u$ to $v$ in the gate graph.
Given this, we have the ground set $U=\{(u_0,v_0),(u_1,v_1),...\}$ where $u_i$ and $v_i$ are each other's 2-hop neighbor.
Then, we also have the edge set $E^\prime$ as the object set $S$ where each edge is assigned with unit cost.
Also, each pair of edge $e_1$ and $e_2$ is associated with single vertex pair $\mathcal{C}(e_1,e_2)$ which can be {\em covered} by them (i.e., $e_1$ and $e_2$ are concatenated to be a shortest path between such pair).
Recall that, our goal is to find a minimal number of edges which are able to form 2-hop shortest paths in the gate graph.
This can be converted to find a subset of edges $E^\ast \subset E^\prime$ such that $\mathcal{C}(E^\ast)=\bigcup_{\{e_i,e_j\}\subseteq E^\ast} \mathcal{C}(e_i, e_j) = U$,
with minimum cost $|E^\ast|$ (i.e., minimum number of edges within $E^\ast$).
Since our problem is the instance of SCP problem which has been proven to be NP-hard, we have:

\bthm
The edge discovery problem on gate graph is NP-hard.
\ethm

\begin{figure}
    \centering
    \mbox{
        \subfigure[Gate Graph] {\epsfig{figure=Figures/gategraphexample.eps,scale=0.6} \label{fig:gategraph}}
        \subfigure[Set-Cover-with-Pairs Setting]{\epsfig{figure=Figures/setcoverpairs.eps,scale=0.6}\label{fig:setcoverwithpairs}}
    }
    \label{scpexample}
    \caption{Example for Gate Graph Simplification Algorithm}
\end{figure}

Let us consider the example in Figure~\ref{scpexample}, where gate graph is shown in Figure~\ref{fig:gategraph}.
Note that, we omit the edge weight in Figure~\ref{fig:gategraph} for simplification, and the shortest paths including 3 vertices are specified in the lower part.
As discussed, each shortest path identified by two endpoints is make up of two edges which serve as objects in the traditional Set-Cover-with-Pairs problem (Definition \ref{scpdef}).
And two endpoints forms a vertex pair included in the ground set.
According to underlying shortest paths, we visualize the edge-pair covering structure in Figure~\ref{fig:setcoverwithpairs}.
There is an edge in the covering structure if and only if two edges form a shortest path in the gate graph.
For instance, there is a shortest path $(v_2,v_8,v_7)$ which is the concatenation of edge $(v_2,v_8)$ and edge $(v_7,v_8)$.
Therefore, we connect $(v_2,v_8)$ and $(v_7,v_8)$ in the edge-pair covering structure.
Also, the shortest path between $v_2$ and $v_7$ can be covered by edge $(v_2,v_3)$ and $(v_3,v_7)$.

\subsection{Gate Graph Simplification with Approximate Distance}
\label{approxgategraph}

In this section, we relax the condition 1) by a user-specified approximate ratio to further sparsify the gate graph.
By adjusting the approximate ratio, we are able to choose a suitable tradeoff between the simplicity and accuracy of shortest distance in the resulting graph.
We first generalize the aforementioned solution based on Set-Cover-with-Pairs (SCP) problem to handle the problem with approximate distance,
then prove that the distance of any pair of vertices in the resulting graph holds user-specified approximation ratio.

Given approximation ratio $\delta$, we wish to exact a subgraph such that the distance of each pair of vertices is not greater than $(1+\delta)$ times of the distance in the original graph.
Intuitively, this flexibility allows us to select fewer edges in the gate graph to satisfied the relaxed condition.
We formally define the problem here:

\bdefin{\bf (Gate Graph Simplification Problem with Approximate Distance)}
\label{ggsimpapproxprob}

Given an undirected graph $G=(V,E)$, its corresponding gate graph $G^\prime=(V^\ast, E^\prime, W)$ and an approximation ratio $\delta$,
we try to find a subgraph $G^\ast=(V^\ast, E^\ast, W^\ast)$ where $E^\ast \subset E^\prime$, $W^\ast \subset W$, and satisfies the following conditions:

\noindent 1) $d_{G^\ast}(u,v) \le (1+\delta) d_{G^\prime} (u,v)$;

\noindent 2) the number of edges $|E^\ast|$ is minimized.

\edefin

Clearly, our problem is the generalization of gate graph simplification with exact distance.
If $\delta=0$, the problem is exactly the previously discussed one.
Interestingly, we observe that our problem with approximate distance can also be converted to the instance of Set-Cover-with-Pairs problem.
Similar to the formulation of gate graph simplification with exact distance,
let us first redefine the concept of {\em edge coverage} in approximate distance setting:

\bdefin{\bf (Edge Coverage with Approximate Distance)}
Let $G=(V,E)$ be original graph, $G^\prime=(V^\ast,E^\prime,W)$ be its corresponding gate graph and user-specific approximation error $\delta$.
There are two edges $e_1=(u,x) \in E^\prime$, $e_2=(y,v) \in E^\prime$, and vertex pair $(u,v)$ where $u$ and $v$ are each other's 2-hop neighbor.
We say $e_1$ and $e_2$ cover vertex pair $(u,v)$ if and only $x$ and $y$ are the same vertex and $w(u,x)+w(y,v) \le (1+\delta) d_G(u,v)$.
\edefin

Let $U=\{(u_0,v_0),(u_1,v_1),...\}$ where $u_i$ and $v_i$ are each other's 2-hop neighbor be the ground set.
Also, the edge set $E^\prime$ is considered as object set $S$ where each edge is assigned with unit cost.
Especially, each pair of edge $e_1$ and $e_2$ is associated with single vertex pair $\mathcal{C}(e_1,e_2)$ which can be {\em covered} by them.
Our goal is to find a subset of edges $E^\ast \subset E^\prime$ such that $\mathcal{C}(E^\ast)=\bigcup_{\{e_i,e_j\}\subseteq E^\ast} \mathcal{C}(e_i, e_j) = U$,
with minimum cost $|E^\ast|$ (i.e., minimum number of edges from $E^\ast$).
The correctness of algorithm is stated in the lemma \ref{correctness}.

\begin{figure}
    \centering
    \mbox{
        \subfigure[Case 1] {\epsfig{figure=Figures/approxproof1.eps,scale=0.6} \label{fig:approxproof1}}
    }
    \mbox{
        \subfigure[Case 2 with direct connection]{\epsfig{figure=Figures/approxproof2.eps,scale=0.6}\label{fig:approxproof2}}
    }
    \mbox{
        \subfigure[Case 2 without direct connection]{\epsfig{figure=Figures/approxproof3.eps,scale=0.6}\label{fig:approxproof3}}
    }
    \label{fig:approxproof}
    \caption{Example for Proof of Lemma \ref{correctness}}
\end{figure}

\blemma
\label{correctness}
Given approximation ratio $\delta$, gate graph $G^\prime=(V^\ast,E^\prime,W)$ and simplified graph $G^\ast=(V^\ast,E^\ast,W^\ast)$,
the distance of each pair of vertices $u$ and $v$ satisfies the property:
\[ d_{G^\ast} \le (1+\delta) d_{G^\prime} (u,v) \]
\elemma

\bproof
Let the path $P=(u, x_1, y_1, ..., v)$ be the shortest path between vertex $u$ and $v$ in the gate graph (as illustrated in Figure \ref{fig:approxproof} wit dashed lines).
Let us suppose $u=y_0$, then the shortest path $P$ contains several 2-hop shortest paths $SP_{y_i,y_{i+1}} = (y_i, x_i, y_{i+1})$.
Based on the Set-Cover-with-Pairs problem formulation, the distance between vertex $y_i$ and its 2-hop neighbors $y_{i+1}$ in the resulting graph is no greater than $(1+\delta)$ times of their shortest distance in the gate graph, i.e., $d_{G^\ast}(y_i,y_{i+1}) \le (1+\delta) d_{G^\prime}(u,v)$.
That is, we guarantee to have one intermediate vertex $x_{i+1}^\prime$ such that $w(y_i,x_{i+1})+w(x_{i+1},y_{i+1}) \le d_{G^\prime}(u,v)$.
In Figure \ref{fig:approxproof}, we show one of the path in the resulting graph connecting vertex $u$ and $v$ (with solid line) passing through such intermediate vertices $x^\prime_i$.
Now, we consider the length of path $P^\prime=(u,x^\prime_1, y_1, ..., v)$ compared to the shortest path $P$ in two cases:
1) the shortest path $P$ contains even number of edges, i.e., $|P|=2k$;
and 2) the shortest path $P$ contains odd number of edges, i.e., $|P|=2k+1$.

\noindent{\bf Case 1: }
Since $|P|=2k$, the shortest path $P$ (with dashed lines) contains $k$ subpaths with 2 edges, such as $(u,x_1,y_1)$, ... $(y_i, x_{i+1}, y_{i+1})$, ..., $(y_{k-1}, x_k, v)$ (as illustrated in Figure \ref{fig:approxproof1}).
According to above discussion, for each subpath $(y_i, x_{i+1}, y_{i+1})$, there must exist one intermediate vertex $x^\prime_{i+1}$ to satisfy the condition:
\[ w(y_i,x_{i+1})+w(x_{i+1},y_{i+1}) \le d_{G^\prime}(u,v) = w(y_i,x_{i+1}^\prime)+w(x_{i+1}^\prime,y_{i+1}) \]
Therefore, assuming $u=y_0$ and $v=y_k$, we have
\beqnarr
d_{G^\ast}(u,v) & = & \sum_{i=0}^{k-1} w(y_i,x_{i+1}^\prime)+w(x_{i+1}^\prime,y_{i+1}) \nonumber \\
 & \le &  \sum_{i=0}^{k-1} w(y_i,x_{i+1})+w(x_{i+1},y_{i+1}) \nonumber \\
 & = & d_{G^\prime} (u,v) \nonumber
\eeqnarr

\noindent{\bf Case2: }
In this case, the shortest path $P$ is made up of a subpath with $2k$ consecutive edges and one additional edge (illustrated in Figure \ref{fig:approxproof2} and Figure \ref{fig:approxproof3}).
For the subgraph with $2k$ edges, we hold the following inequality based on case 1:
\beqnarr
d_{G^\ast}(u,y_k) \le d_{G^\prime}(u,y_k) \nonumber
\eeqnarr
Here, we consider the last edge of shortest path $P$ in two different cases: 1) last vertex $y_k$ directly connects to destination $v$ (shown in Figure \ref{fig:approxproof2});
2) $y_k$ does not directly connect to $v$ (shown in Figure \ref{fig:approxproof3}).

For the first case, we easily have following:
\beqnarr
d_{G^\ast}(u,v) & = & d_{G^\ast}(u,y_k) + w(y_k,v) \nonumber \\
& \le &  (1+\delta) d_{G^\prime}(u,y_k) + w(y_k,v) \nonumber \\
& \le &  (1+\delta) (d_{G^\prime}(u,y_k) + w(y_k,v)) \nonumber \\
& = & (1+\delta) d_{G^\prime}(u,v) \nonumber
\eeqnarr

For the second case, consider $v$'s 2-hop neighbor which lies in the shortest path $P$, i.e., the vertex $x_k$ in Figure \ref{fig:approxproof3}.
We first prove that $x_k^\prime$ is $v$'s 2-hop neighbor, then construct an alternate path to connect $x^\prime_k$ and $v$.

According to Set-Cover-with-Pair problem formulation, there should exist one vertex $y^\prime_k$ having $w(x_k,y^\prime_k) + w(y^\prime_k,v) \le (1+\delta) d_{G^\prime}(x_k,v)$.

\eproof
}

\section{Experimental Evaluation}
\label{experiments}

In this section, we empirically study the performance of our approaches on both real and synthetic datasets.
Specifically, we compare two methods in the experiments:
1) {\bf FS}, which corresponds to the approach utilizing adaptive sampling~\cite{Tao11} for gate vertices discovery;
2) {\bf SC}, which corresponds to the approach using set cover framework for gate vertices discovery (Subsection~\ref{setcovergate}).
Here, we are interested in understanding how many vertices can be reduced by the gate-vertex set and how many edges are needed in the gate graph,
and how they are affected by the locality parameter $\epsilon$?
In each experiment, we measure the number of gate vertices and the number of edges in gate graph, and the running time of algorithms.
To gain a better understanding of experimental results, we also report two important graph measures: diameter (refer to as {\bf Diam.}) and average value of pairwise shortest distances (refer to as {\bf Avg.Dist}), for each graph.
We implemented our algorithms in C++ and Standard Template Library (STL).
All experiments were conducted on a 2.8GHz Intel Xeon CPU and 12.0GB RAM running Linux 2.6.



\begin{table}
\begin{center}
{\small


\begin{tabular}{|l|c|c|c|c|}
\hline
Dataset &  \#V  & \#E   & Dia. & Avg.Dist \\ \hline
CA-GrQc &   5242    &   28980    &   17  &   6.1 \\ \hline
CA-HepTh    &   9877    &   51971    &   18  &   6   \\ \hline
Wiki-Vote   &   7115    &   103689   &   7   &   3.3 \\ \hline
P2PG08  &   6301    &   20777   &     9  &    4.6    \\ \hline
P2PG09  &   8114    &   26013   &   9   &   4.8    \\ \hline
P2PG30  &   36682   &   88328   &   11  &   5.7     \\ \hline
P2PG31  &   62586   &   147892   &   11  &   5.9     \\ \hline
\end{tabular}
}
\end{center}
\vspace{-2.0ex}
\caption{Real Datasets}
\label{tab:realdata}
\vspace*{-3.0ex}
\end{table}


\subsection{Real Data}
\label{realdata}

In this subsection, we collect 7 real-world datasets listed in Table~\ref{tab:realdata} to validate the performance of our approaches.
Among them, CA-GrQc and CA-HepTh are collaboration networks from arXiv describing scientific collaboration relationships between authors in General Relativity and Quantum Cosmology field, and in High Energy Physics field, respectively.
Moreover, P2PG08, P2PG09, P2PG30 and P2PG31 are $4$ snapshots of the Gnutella peer-to-peer file sharing network collected in August and September 2002, respectively.
Wiki-Vote describes the relationships between users and their related discussion from the inception of Wikipedia until January 2008.
All datasets are publicly available at Stanford Large Network Dataset Collection~\footnote{http://snap.stanford.edu/data}.

\begin{figure*}[tbp]
\resizebox{\textwidth}{!} {
    \begin{minipage}[c]{0.33\textwidth}
    \centering
    \includegraphics[height=1.5in]{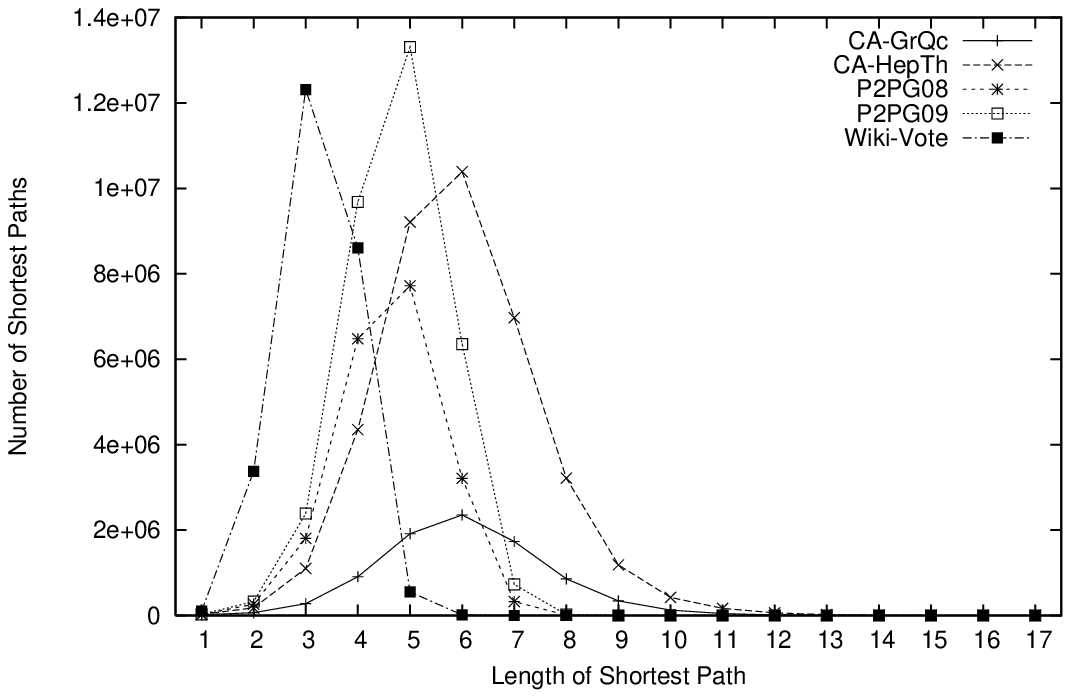}
    \caption {{\small  Distance Distrib.(Real Graph)}}
    \label{fig:realdist}
    \end{minipage}
    \hspace{0.02cm}

    \begin{minipage}[c]{0.33\textwidth}
    \centering
    \includegraphics[height=1.5in]{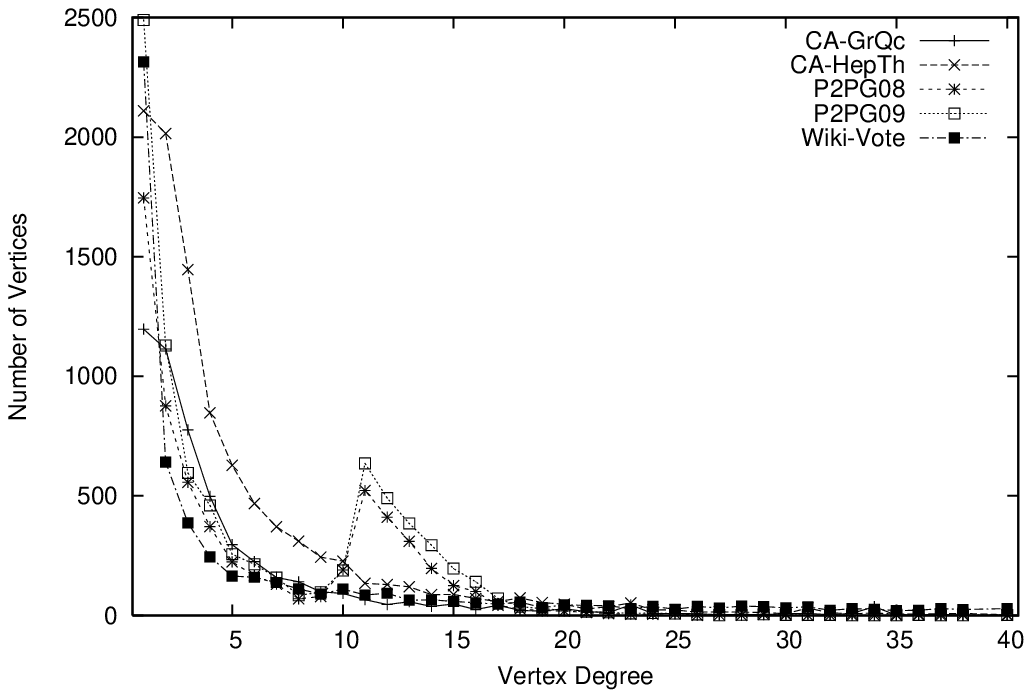}
    \caption {{\small Degree Distrib. (Real Graph)}}
    \label{fig:realdeg}
    \end{minipage}
    \hspace{0.02cm}

    \begin{minipage}[c]{0.33\textwidth}
    \centering
    \includegraphics[height=1.5in]{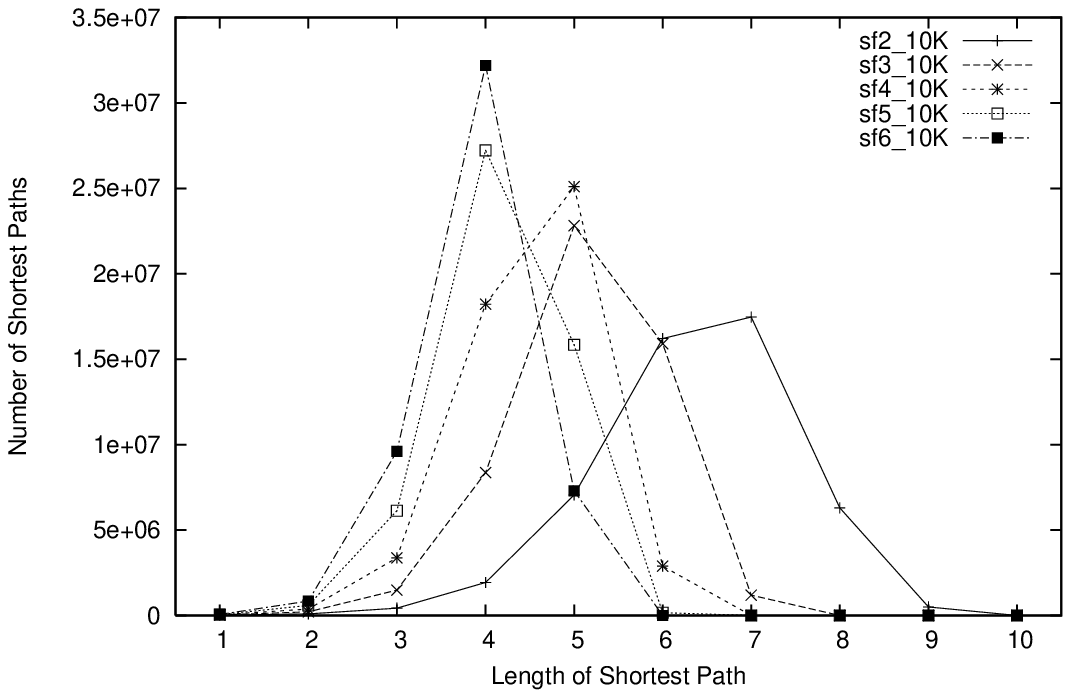}
    \caption {{\small  Distance Distrib. (Scale-free Graph)}}
    \label{fig:pl10kdist}
    \end{minipage}
    \hspace{0.02cm}

}
\\
\resizebox{\textwidth}{!} {
    \hspace{0.02cm}

    \begin{minipage}[c]{0.33\textwidth}
    \centering
    \includegraphics[height=1.5in]{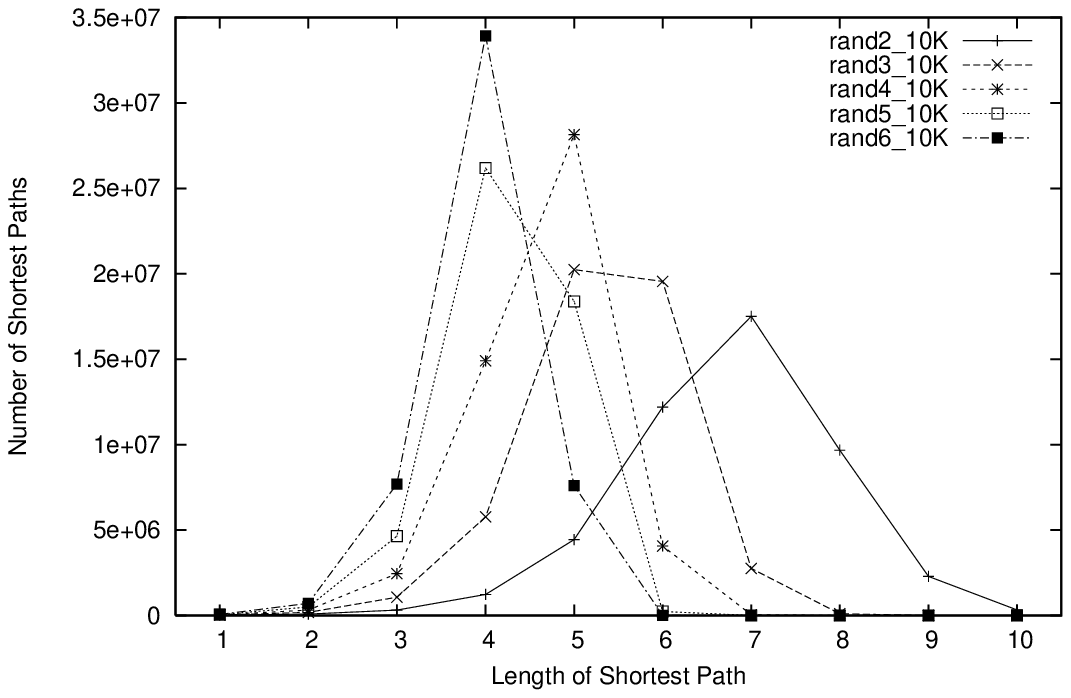}
    \caption {{\small Distance Distrib. (Erd\"{o}s-R\'{e}nyi Graph)}}
    \label{fig:rand10kdist}
    \end{minipage}
    \hspace{0.02cm}

    \begin{minipage}[c]{0.33\textwidth}
    \centering
    \includegraphics[height=1.5in]{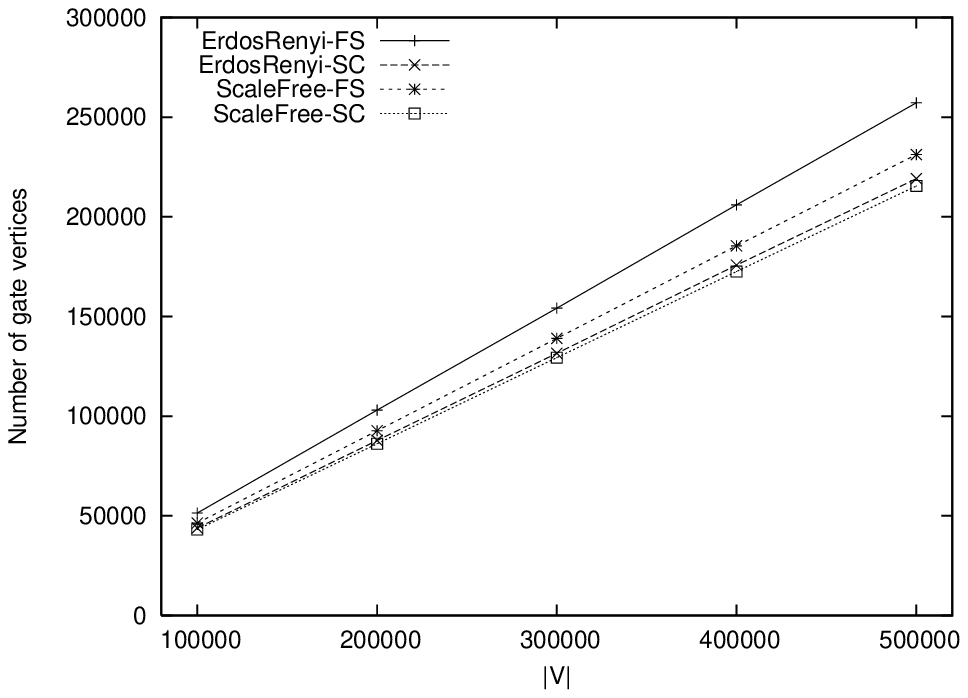}
    \caption {{\small Gate-Vertex Size (Large Rand Graph)}}
    \label{fig:ggvertexrandom}
    \end{minipage}
    \hspace{0.02cm}

    \begin{minipage}[c]{0.33\textwidth}
    \centering
    \includegraphics[height=1.5in]{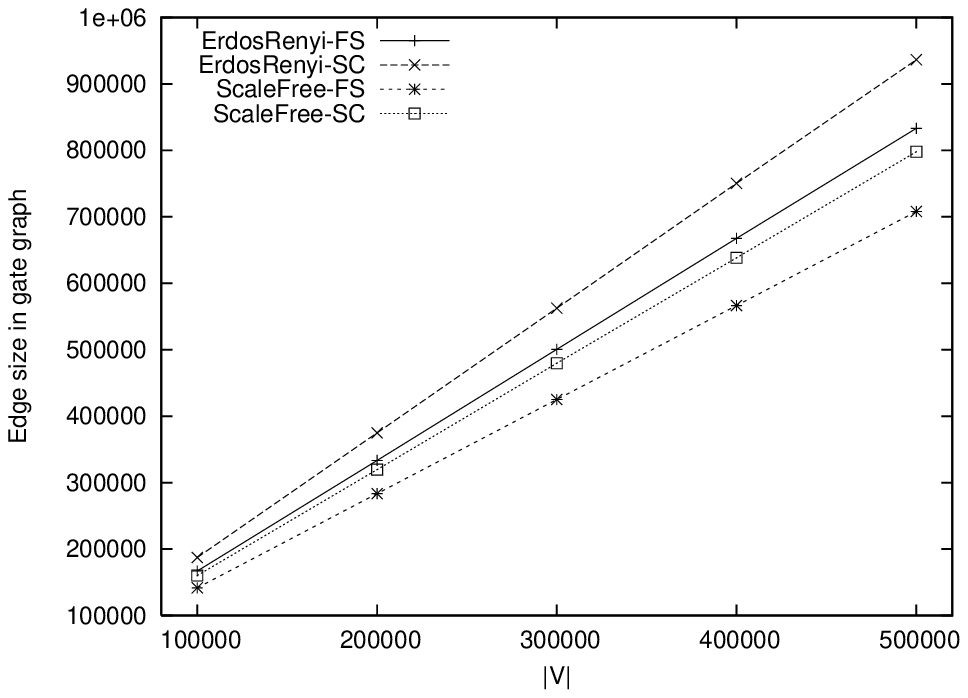}
    \captionof{figure} {{\small Edge Size in Gate Graph (Large Rand Graph)}}
    \label{fig:ggedgerandom}
    \end{minipage}
    \hspace{0.02cm}
}
\end{figure*}

Table \ref{tab:realdataresult} reports the size of gate-vertex set and the number of edges in gate graphs by varying locality parameter $\epsilon$ from $3$ to $6$.
Their corresponding shortest distance distribution and vertex degree distribution are shown in Figure~\ref{fig:realdist} and Figure~\ref{fig:realdeg}, respectively.
Since the distances and vertex degrees of P2PG30 and P2PG31 have similar distribution with that of P2PG08 and P2PG09,
and their large values would affect other datasets' distribution visualization, we omit them in both figures.
We make the following observations:

\noindent{\bf Size of Gate-Vertex Set:}
Table \ref{tab:realdataresult} shows that the sizes of gate-vertex set discovered by both FS and SC are consistently smaller than that of original graphs.
Among them, SC always obtains the better results, which are on average approximately $76\%$, $65\%$, $63\%$ and $56\%$ of the one from FS with $\epsilon$ ranging from $3$ to $6$.
For SC approach, the size of gate-vertex set by SC is on average around $26\%$, $21\%$, $27\%$ and $24\%$ of the corresponding original graph when $\epsilon$ varies from $3$ to $6$.
We also observe that, as locality parameter $\epsilon$ increases, the number of gate vertices discovered by SC is gradually reduced.
Particularly, reduction ratios of CA-GrQc, CA-HepTh and Wiki-Vote are consistently better than that of P2P08, P2P09, P2P30 and P2P31.
In Figure~\ref{fig:realdeg}, CA-GrQc, CA-HepTh and Wiki-Vote seem to fit the power-law degree distribution very well, while there are a significant portion of vertices with degree ranging from $10$ to $15$ in P2P08, P2P09, P2P30 and P2P31.
In other words, there exists a small portion of vertices with high degree potentially serving as the intermediate connectors for traffics between a large portion of vertex pairs in CA-GrQc, CA-HepTh and Wiki-Vote.
By SC's gate vertices discovery method using set cover framework, those vertices can be selected as gate vertices and thus dramatically simplify original graphs.
However, for file-sharing network, a relatively large number of vertices with high connectivity potentially leads to larger size of gate-vertex set by the same selection principle.
From the perspective of application domains, the results of SC on three social networks (i.e., CA-GrQc, CA-HepTh and Wiki-Vote) suggest a small highway structure capturing major non-local communications in the network.
Interestingly, the consistent decreasing trends regarding the size of gate-vertex set with increasing $\epsilon$ are not observed in the results of FS on P2P30 and P2P31.
Since adaptive sampling approach follows the spirit of greedy algorithm -  choosing each gate vertex only based on local information,
the mis-selection of gate vertices at earlier stages probably leads to significant increase of gate vertices at later stages.
In other words, some important vertices selected as gate vertices in the procedure with small $\epsilon$ might be missed in the procedure with larger $\epsilon$.
Therefore, it is reasonable to observe that the number of gate vertices discovered by FS unexpectedly becomes larger when $\epsilon$ increases.

\comment{
\noindent{\bf 1.} The number of gate-vertex set is much smaller than that of the original graph.
On average, the number of gate-vertex set is only $26.7\%$, $21.5\%$, $27.8\%$, and $24.7\%$ of the corresponding original graph when $\epsilon$ varies from $2$ to $5$.
In addition, as the locality parameter $\epsilon$ increases, the number of discovered gate-vertex is reduced.
Particularly, the reduction ratios of CA-GrQc, CA-HepTh and Wiki-Vote are consistently better than that of P2P08, P2P09, P2P30 and P2P31 in terms of the number of gate vertices.
In Figure~\ref{fig:realdeg}, CA-GrQc, CA-HepTh and Wiki-Vote seem to fit the power-law degree distribution very well, while there are a significant portion of vertices with degree ranging from $10$ to $15$ in P2P08, P2P09, P2P30 and P2P31.
In other words, there exists a small portion of vertices with high degree potentially serving as the intermediate connectors for traffics between a large portion of vertex pairs in CA-GrQc, CA-HepTh and Wiki-Vote.
According to our gate-vertex algorithm based on set cover framework, those vertices can be selected as gate-vertex and thus can help dramatically reduce the number of vertices in the original graph.
However, for file-sharing network, there are a relatively large number of vertices which have relatively high connectivity and these vertices can potentially increase the number of gate vertices.
From the perspective of application domains, our result on the three social networks (i.e., CA-GrQc, CA-HepTh and Wiki-Vote) suggest a small highway structure capturing major non-local communications in the network.
}

\noindent{\bf Edge Size of Gate Graph:}
The number of edges in original graphs are significantly reduced by SC on three datasets CA-GrQc, CA-HepTh and Wiki-Vote.
Especially, on average, the number of edges in gate graphs generated by SC are  $6.5$, $6$, $6.3$ and $6$ times smaller than that of original graphs for $\epsilon$ to be $3$, $4$, $5$ and $6$.
Besides that, SC still outperforms FS on those datasets, such that the number of edges in gate graphs by SC are on average about $49\%$, $51\%$, $53\%$ and $48\%$ of the one from gate graph by FS ranging $\epsilon$ from $3$ to $6$.
Interestingly, as $\epsilon$ becomes larger, the number of edges in gate graphs generated by SC increases on CA-GrQc and CA-HepTh.
The reason is, in order to guarantee that shortest paths between all non-local vertex pairs can be recovered utilizing fewer gate vertices,
more edges are needed to build stronger connections among gate vertices.
However, the number of edges in gate graphs generated by FS on CA-GrQc and CA-HepTh becomes smaller when $\epsilon$ increases.
This demonstrates the effectiveness of edge sparsification algorithm for pruning redundant edges, since some of gate vertices discovered by FS are non-essential and are not necessarily to be connected to its $\epsilon$ neighbors.
For other four datasets (P2P08, P2P09, P2P30 and P2P31), gate graphs generated by FS from those datasets contains fewer edges compared to the one of SC.
Overall, they are on average about $1.3$, $1.1$, $1.4$ and $1.8$ times smaller than that of SC varying $\epsilon$ from $3$ to $6$.
Also, as $\epsilon$ increases, the number of edges in gate graphs generated by both approaches on those four datasets increases.
This is consistent with our earlier discussion that in these graphs, their interactions seem to be more random and the relatively large number of vertices with degrees between $10$ to $15$ may increase their chance to connect to other vertices with local walks.

\comment{
\noindent{\bf 2.} In terms of the number of edges in the gate graphs, the three datasets with better reduction ratio on the number of vertices also perform better than the other four.
Especially, in datasets CA-GrQc, CA-HepTh and Wiki-Vote, it is interesting to see that the number of edges decreases as $\epsilon$ increases.
On average, the number of edges in these gate graphs are only around $30.9\%$, $22.5\%$, $18.3\%$ and $17\%$ of the number of edges in the original graphs for $\epsilon$ to be $2$, $3$, $4$ and $5$, respectively.
In addition, the average vertex degree in these gate graphs are rather close to the vertex degree in the original graphs.
However, for the other four datasets (P2P08, P2P09, P2P30 and P2P31), the number of edges in their gate graphs is on average approximately $1.1$, $1.4$, $2.0$ and $2.6$ times greater than the one of original graphs with locality parameter $\epsilon$ to be $2$, $3$, $4$, and $5$.
This is also consistent with our earlier discussion that in these graphs, their interactions seem to be more random and the relatively large number of vertices with degrees between $10$ to $15$ may increase their chance to connect to other vertices with local walks.
}

\noindent{\bf Running Time: }
We take $\epsilon=3$ as an example.
The running time of FS for all $7$ datasets are $65$ms, $132$ms, $3$s, $127$ms, $158$ms, $447$ms and $811$ms.
The running time of SC are $23$s,  $53$s, $1166$s, $183$s, $293$s, $279$s and $661$s for CA-GrQc, CA-HepTh, Wiki-Vote, P2P08, P2P09, P2P30 and P2P31, respectively.
As locality parameter $\epsilon$ increases, the computational cost of both approaches become larger,
because more vertex pairs should be considered in SC and more vertices would be traversed in FS.
The average running time of SC on $\epsilon=5$ can cost up to a few hours, which is around $100$ times slower than that of FS.
Indeed, the selection between FS and SC is a trade-off between reduction ratio and efficiency.
In general, we can see that with rather smaller $\epsilon$ ($2$ or $3$),
the vertex reduction by SC is quite significant which is also much better than that of FS,
and their running time are reasonable in practice.
In contrast to FS, the size of gate-vertex set discovered by SC is guaranteed to hold logarithmic approximation bound.
Therefore, we would say SC with smaller $\epsilon$ is applicable in most of applications.

\comment{
For SC utilizing set cover framework, when the value of $\epsilon$ increases, more vertex pairs with distance $\epsilon+1$ needs to be considered, therefore the computational cost of collecting those vertex pairs (i.e., serve as ground set) and their candidate sets (the construction of set-cover instance) become dominant cost.
The computational cost of FS becomes larger as $\epsilon$ increases because more vertices need to be traversed during the step of performing BFS.
The average running time of SC on $\epsilon=5$ can cost up to a few hours, which is around $10$ times slower than that of FS.
However, in general we can see that with rather smaller $\epsilon$ ($2$ or $3$),
the vertex reduction by SC is already quite significant which is also much better than that of FS,
its running time is rather reasonable in practice.
Thus...
 and the improvement using bigger $\epsilon$ does not necessarily payoff (as it is more computationally expensive).
}

\comment{
\noindent{\bf 3.} For the running time, take $\epsilon=2$ as an example.
The running time for those seven datasets are $48$s,  $52$s, $400$s, $650$s, $277$s, $662$s and $2030$s for CA-GrQc, CA-HepTh, P2P08, P2P09, P2P30, P2P31, and Wiki-Vote.
When the value of $\epsilon$ increases, more vertex pairs with distance $\epsilon+1$ needs to be considered for set cover framework, therefore the computational cost of collecting those vertex pairs (i.e., serve as ground set) and their candidate sets (the construction of set-cover instance) become dominant cost.
The average running time on $\epsilon=5$ can cost up to a few hours.
However, in general we can see that with rather smaller $\epsilon$ ($2$ or $3$), the vertex reduction is already quite significant and the improvement using bigger $\epsilon$ does not necessarily payoff (as it is more computationally expensive).
}

\begin{table*}[!htbp]
\resizebox{\textwidth}{!} {\small
\begin{tabular}{|l|r|r|r|r|r|r|r|r|r|r|r|r|r|r|r|r|}
\hline
\multirow{3}{*}{Dataset}
& \multicolumn{4}{c|}{$\epsilon=3$} & \multicolumn{4}{c|}{$\epsilon=4$} & \multicolumn{4}{c|}{$\epsilon=5$}
& \multicolumn{4}{c|}{$\epsilon=6$} \\ 
 \cline{2-17}
 &
\multicolumn{2}{c|}{$\#V$} & \multicolumn{2}{c|}{$\#E$} & \multicolumn{2}{c|}{$\#V$} & \multicolumn{2}{c|}{$\#E$}
 & \multicolumn{2}{c|}{$\#V$} & \multicolumn{2}{c|}{$\#E$} &
\multicolumn{2}{c|}{$\#V$} & \multicolumn{2}{c|}{$\#E$} 
\\
\cline{2-17}
&
\multicolumn{1}{c|}{FS} & \multicolumn{1}{c|}{SC} & \multicolumn{1}{c|}{FS} & \multicolumn{1}{c|}{SC}
 & \multicolumn{1}{c|}{FS} & \multicolumn{1}{c|}{SC} &
\multicolumn{1}{c|}{FS} & \multicolumn{1}{c|}{SC} &
\multicolumn{1}{c|}{FS} & \multicolumn{1}{c|}{SC} & \multicolumn{1}{c|}{FS} & \multicolumn{1}{c|}{SC}
 & \multicolumn{1}{c|}{FS} & \multicolumn{1}{c|}{SC} &
\multicolumn{1}{c|}{FS} & \multicolumn{1}{c|}{SC}
\\ \hline

CA-GrQc &   2836    &   869 &   9266    &   2655    &   1625    &   655 &   6848    &   2933    &   1116    &   567 &   5580    &   2984    &   908 &   500 &   5192    &   2858    \\ \hline
CA-HepTh    &   5131    &   2208    &   15831   &   7674    &   3381    &   1669    &   14921   &   10241   &   2525    &   1364    &   14316   &   11249   &   2134    &   1157    &   14476   &   11456   \\ \hline
Wiki-Vote   &   2564    &   1598    &   84607   &   59132   &   2457    &   879 &   85051   &   34736   &   2236    &   584 &   83681   &   22652   &   2964    &   571 &   193913  &   19571   \\ \hline

P2P08   &   2359    &   2340    &   10892   &   10738   &   2313    &   1920    &   23787   &   25406   &   2082    &   1584    &   26497   &   40218   &   2043    &   1095    &   28002   &   49139   \\ \hline
P2P09   &   2930    &   2904    &   13633   &   13394   &   2874    &   2474    &   30219   &   32976   &   2643    &   2047    &   34870   &   53851   &   2556    &   1530    &   37022   &   75113   \\ \hline
P2P30   &   9688    &   9627    &   37708  &   36708   &   10874   &   8820    &   98194   &   92820   &   8713    &   7551    &   108720  &   151677  &   8914    &   6845    &   127565  &   232971  \\ \hline
P2P31   &   16493   &   16394   &   64624  &   146765  &   18248   &   14996   &   161883  &   155099  &   14745   &   12847   &   182599  &   256578  &   14895   &   11738   &   215742  &   383725   \\ \hline
%

\end{tabular}
}
\vspace{-2.0ex}
\caption{Sizes of Simplified Graph on Real Datasets}
\label{tab:realdataresult}
\vspace*{-2.0ex}
\end{table*}

\begin{table*}[!htbp]
\resizebox{\textwidth}{!}
{\small
\begin{tabular}{|l|r|r|r|r|r|r|r|r|r|r|r|r|r|r|r|r|r|r|r|r|}
\hline \multirow{3}{*}{Dataset} & 
\multicolumn{4}{c|}{$\epsilon=3$} & \multicolumn{4}{c|}{$\epsilon=4$} & \multicolumn{4}{c|}{$\epsilon=5$}
& \multicolumn{4}{c|}{$\epsilon=6$} & \multicolumn{4}{c|}{$\epsilon=7$}  \\
 \cline{2-21}
 &
\multicolumn{2}{c|}{$\#V$} & \multicolumn{2}{c|}{$\#E$} & \multicolumn{2}{c|}{$\#V$} & \multicolumn{2}{c|}{$\#E$}
 & \multicolumn{2}{c|}{$\#V$} & \multicolumn{2}{c|}{$\#E$} &
\multicolumn{2}{c|}{$\#V$} & \multicolumn{2}{c|}{$\#E$}  & \multicolumn{2}{c|}{$\#V$} & \multicolumn{2}{c|}{$\#E$}
\\
\cline{2-21}
&
\multicolumn{1}{c|}{FS} & \multicolumn{1}{c|}{SC} & \multicolumn{1}{c|}{FS} & \multicolumn{1}{c|}{SC}
 & \multicolumn{1}{c|}{FS} & \multicolumn{1}{c|}{SC} &
\multicolumn{1}{c|}{FS} & \multicolumn{1}{c|}{SC} &
\multicolumn{1}{c|}{FS} & \multicolumn{1}{c|}{SC} & \multicolumn{1}{c|}{FS} & \multicolumn{1}{c|}{SC}
 & \multicolumn{1}{c|}{FS} & \multicolumn{1}{c|}{SC} &
\multicolumn{1}{c|}{FS} & \multicolumn{1}{c|}{SC}
 & \multicolumn{1}{c|}{FS} & \multicolumn{1}{c|}{SC} &
\multicolumn{1}{c|}{FS} & \multicolumn{1}{c|}{SC}
\\ \hline

sf2\_10K    &   5781    &   5499    &   10559   &   7931    &   4648    &   4273    &   14067   &   15847   &   4057    &   3523    &   17175   &   21206   &   3754    &   3015    &   20258   &   26123   &   3495    &   2547    &   22127   &   32003   \\ \hline
sf3\_10K    &   6627    &   6203    &   18824   &   14926   &   5673    &   5208    &   29760   &   36385   &   5239    &   4437    &   39216   &   55411   &   4920    &   3457    &   46244   &   86414   &   4655    &   1264    &   50130   &   64944   \\ \hline
sf4\_10K    &   7190    &   6661    &   27721   &   22485   &   6406    &   5797    &   47198   &   61880   &   5992    &   4763    &   63681   &   103913  &   5737    &   1593    &   72145   &   124394  &   5211    &   800 &   92447   &   56418   \\ \hline
sf5\_10K    &   7647    &   7012    &   37339   &   30257   &   6968    &   6229    &   64952   &   89829   &   6641    &   4220    &   86701   &   204039  &   6315    &   802 &   96952   &   32989   &   5876    &   0   &   308613  &   0   \\ \hline
sf6\_10K    &   7912    &   7278    &   46570   &   38437   &   7315    &   6539    &   84063   &   120231  &   7035    &   2266    &   110166  &   218814  &   6686    &   789 &   123614  &   59338   &   5969    &   0   &   863493  &   0   \\ \hline

\end{tabular}
}
\vspace{-2.0ex}
\caption{Sizes of Simplified Graph on Scale-free Graphs}
\label{tab:scalefreeresult}
\end{table*}

\begin{table*}[!htbp]
\resizebox{\textwidth}{!}
{
\begin{tabular}{|l|r|r|r|r|r|r|r|r|r|r|r|r|r|r|r|r|r|r|r|r|}
\hline \multirow{3}{*}{Dataset} & 
\multicolumn{4}{c|}{$\epsilon=3$} & \multicolumn{4}{c|}{$\epsilon=4$} & \multicolumn{4}{c|}{$\epsilon=5$}
& \multicolumn{4}{c|}{$\epsilon=6$} & \multicolumn{4}{c|}{$\epsilon=7$}  \\
 \cline{2-21}
 &
\multicolumn{2}{c|}{$\#V$} & \multicolumn{2}{c|}{$\#E$} & \multicolumn{2}{c|}{$\#V$} & \multicolumn{2}{c|}{$\#E$}
 & \multicolumn{2}{c|}{$\#V$} & \multicolumn{2}{c|}{$\#E$} &
\multicolumn{2}{c|}{$\#V$} & \multicolumn{2}{c|}{$\#E$}  & \multicolumn{2}{c|}{$\#V$} & \multicolumn{2}{c|}{$\#E$}
\\
\cline{2-21}
&
\multicolumn{1}{c|}{FS} & \multicolumn{1}{c|}{SC} & \multicolumn{1}{c|}{FS} & \multicolumn{1}{c|}{SC}
 & \multicolumn{1}{c|}{FS} & \multicolumn{1}{c|}{SC} &
\multicolumn{1}{c|}{FS} & \multicolumn{1}{c|}{SC} &
\multicolumn{1}{c|}{FS} & \multicolumn{1}{c|}{SC} & \multicolumn{1}{c|}{FS} & \multicolumn{1}{c|}{SC}
 & \multicolumn{1}{c|}{FS} & \multicolumn{1}{c|}{SC} &
\multicolumn{1}{c|}{FS} & \multicolumn{1}{c|}{SC}
 & \multicolumn{1}{c|}{FS} & \multicolumn{1}{c|}{SC} &
\multicolumn{1}{c|}{FS} & \multicolumn{1}{c|}{SC}
\\ \hline

rand2\_10K  &   6243    &   5442    &   11523   &   7303    &   5158    &   4400    &   16627   &   18510   &   4709    &   3793    &   20557   &   27434   &   4386    &   3430    &   24200   &   36691   &   4165    &   3053    &   27296   &   47571   \\ \hline
rand3\_10K  &   7131    &   6340    &   20095   &   13863   &   6351    &   5449    &   33016   &   42290   &   5925    &   4873    &   43733   &   67867   &   5688    &   4071    &   52679   &   117127  &   5489    &   1933    &   57828   &   231900  \\ \hline
rand4\_10K  &   7727    &   6910    &   29142   &   21027   &   7108    &   6134    &   51635   &   72715   &   6801    &   5378    &   69228   &   130750  &   6608    &   2122    &   80754   &   305901  &   6348    &   830 &   90710   &   70541   \\ \hline
rand5\_10K  &   8157    &   7294    &   38769   &   28824   &   7629    &   6572    &   71324   &   106953  &   7367    &   4654    &   96131   &   287915  &   7170    &   847 &   108997  &   92646   &   6618    &   9   &   147143  &   36  \\ \hline
rand6\_10K  &   8452    &   7560    &   48476   &   36858   &   8005    &   6958    &   91149   &   142715  &   7797    &   2190    &   121549  &   278487  &   7607    &   802 &   134621  &   88650   &   6533    &   0   &   252001  &   0   \\ \hline

\end{tabular}
}
\vspace{-2.0ex}
\caption{Sizes of Simplified Graph on Erd\"{o}s-R\'{e}nyi Random Graphs} \label{tab:randomresult}
\end{table*}

\subsection{Synthetic Data}
\label{syndata}

In the following, we study two approaches on Scale-Free and Erd\"{o}s-R\'{e}nyi random graphs.

\noindent{\bf Scale-Free Random Graph: }
In this experiment, we generated a set of scale-free random graphs such that vertex degree follows power-law distribution using a publicly available graph generator~\footnote{http://pywebgraph.sourceforge.net}.
The number of vertices in those graphs are $10K$, and their edge density (i.e., $|E|/|V|$) ranges from $2$ to $6$.
The diameter of those graphs are $10$, $8$, $7$, $6$, $6$, and their average pairwise distance are $6.4$, $5.1$, $4.5$, $4.2$ and $3.9$.

We can see from Table~\ref{tab:scalefreeresult}, when locality parameter $\epsilon$ increases, the size of gate-vertex set discovered by both approaches consistently decreases for graphs with different edge density.
Similar to the observation in the real-world datasets, SC always achieves better results than FS in terms of the number of gate vertices.
Overall, the size of gate-vertex set discovered by SC is on average around $93\%$, $90\%$, $69\%$, $40\%$ and $28\%$ of the one of FS with $\epsilon$ from $3$ to $7$.
In addition, as edge density increases, when locality parameter $\epsilon$ less than $Avg.Dist$, more gate vertices are discovered by both FS and SC in denser graphs.
For denser graphs, since graph diameter becomes smaller, much more vertex pairs with distance $\epsilon$ need to be covered compared to sparse graphs (see Figure~\ref{fig:pl10kdist}).
Therefore, more gate vertices are required to serve as intermediate hops for any vertex pair.
When $\epsilon$ is greater than $Avg.Dist$,
the number of gate vertices discovered by SC is dramatically reduced since fewer vertex pairs need to be processed in set cover framework (e.g., sf5\_10K and sf6\_10K with $\epsilon \ge 6$).
However, this phenomena is not observed in the results of FS, i.e., their gate-vertex sets are reduced very slowly.
Even when $\epsilon$ is greater than diameter, no gate vertex is actually needed while FS still discovers lots of gate vertices (e.g., sf5\_10K and sf6\_10K with $\epsilon=7$).
In terms of the number of edges in gate graphs, FS performs slightly better than SC.
When locality parameter $\epsilon$ is less than $Avg.Dist$, the results on both FS and SC consistently increase following the opposite trend of the number of gate vertices.
With the increase of $\epsilon$, fewer gate vertices are discovered and more internal connections within gate graph should be built to guarantee that there is a shortest path between any pair of gate vertices.

\comment{
In terms of the number of edges in gate graph,
though fewer gate vertices are discovered by SC, the number of edges in their gate graphs are slightly larger than that of FS.
Overall, they are on average approximately $1.3$, $1.7$, $1.2$ and $1.1$ times greater than the one from FS with $\epsilon$ to be $4$, $5$, $6$ and $7$.
As expected, some gate vertices discovered by FS are non-essential and large portion of redundant edges connecting them in local-gate graph are pruned by our edge sparsification procedure.
As $\epsilon$ increases, the results of FS always become larger.
The increasing trend is observed in the results of SC as well only when $\epsilon$ is less than $Avg.Dist$.
With the increase of $\epsilon$, fewer gate vertices are discovered and more internal connections within gate graph should be build to guarantee that there is a shortest path between any pair of gate vertices.
}

In terms of running time, when edge density increases, running times of both approaches are consistently increased.
Taking $\epsilon=3$ as example, running time of SC for scale-free graphs with density from $2$ to $6$ are $6$s, $49$s, $158$s, $531$s and $1067$s, respectively.
The running time of FS for those graphs are $85$ms, $170$ms, $232$ms, $344$ms and $471$ms, respectively.
As $\epsilon$ becomes larger, longer running time is expected due to more vertices will be visited in both SC (i.e., procedure {\em BFSSetCoverConstruction}) and FS (i.e., breath-first-search).
When $\epsilon=6$, the running time of SC is on average around $30$ times longer than that of $\epsilon=3$, ranging from $430$s to $6617$s.
Also, running time of FS with $\epsilon=6$ is significantly increased which varies from $18$s to $1966$s.
As $\epsilon$ increases, the efficiency advantage of FS over SC is dramatically reduced.
This is caused by the explosive increase on the running time of FS's edge sparsification, since the number of edges in local-gate graph of SC is much smaller than that of FS when $\epsilon \ge 6$.

\comment{
Similar to the observation in the real-world datasets, when the locality parameter $\epsilon$ increases, the number of gate-vertex consistently decreases for random graphs with different edge density.
Overall, the number of vertices in original graphs are reduced by $35\%$, $44\%$, $58\%$ for $\epsilon=2$, $\epsilon=3$ and $\epsilon=4$ (we omit the reduction ratios with $\epsilon=4$ and $\epsilon=5$ since most distances become local distances).
In addition, as edge density increases, when the locality parameter $\epsilon$ less than $Avg.Dist$, more gate-vertex are discovered in denser graphs.
For denser graphs, since graph diameter becomes smaller, much more vertex pairs with distance $\epsilon+1$ are generated and needs to be covered compared to sparse graphs (illustrated in Figure~\ref{fig:pl10kdist}).
Therefore, more gate-vertex are required to serve as intermediate hops for any vertex pair.
In terms of the number of edges in gate graph, when locality parameter is less than $Avg.Dist$, they consistently increase following the similar trend of the number of gate-vertex.
With the increase of $\epsilon$, fewer gate-vertex are discovered and more internal connections within gate graph should be build to guarantee that there is a shortest path between any pair of gate-vertex.

In terms of running time, an variant of breath-first-search is introduced to collect information for greedy set cover, and this collecting procedure dominates total computational cost.
Furthermore, breath-first-search takes longer time on dense graph rather than sparse graph.
Therefore, as the edge density increase, the running time is consistently increased.
Taking $\epsilon=2$ as example, the running time for scale-free graphs with density from $2$ to $6$ are $7$s, $50$s, $160$s, $536$s and $1065$s, respectively.
As locality parameter $\epsilon$ becomes larger, longer running time is expected due to larger number of vertices to be visited by BFS.
When $\epsilon=5$, the running time is on average around $30$ times longer than that of $\epsilon=2$, ranging from $503$s to $6665$s.
}

\noindent{\bf Erd\"{o}s-R\'{e}nyi Random Graph: }
In this experiment, we generate a set of random graphs based on Erd\"{o}s-R\'{e}nyi model,
with the edge density from $2$ to $6$, while keeping the number of vertices at $10K$.
The diameter of those random graphs are $14$, $10$, $8$, $7$ and $6$, respectively.
Also, their corresponding average pairwise distance are $6.8$, $5.3$, $4.7$, $4.3$ and $4.0$, respectively.
Their shortest distance distribution is presented in Figure~\ref{fig:rand10kdist}.
By varying $\epsilon$ from $3$ to $7$, Table~\ref{tab:randomresult} shows the number of vertices and the number of edges in simplified graphs with respect to original graphs with different edge density.
The observations for both approaches SC and FS on scale-free graphs are still hold on Erd\"{o}s-R\'{e}nyi random graphs.
Overall, the sizes of gate-vertex set discovered by SC are on average approximately $88\%$, $86\%$, $66\%$, $41\%$ and $30\%$ of the one from FS with $\epsilon$ from $3$ to $7$.
In terms of the number edges in gate graphs, FS achieves slightly better results than SC.
When $\epsilon$ is no less than $4$, the number of edges in gate graphs with different edge density by SC are on average $1.3$, $2$, $1.8$ and $1.6$ times greater than that of FS, respectively.
Given the same $\epsilon$ and edge density, the size of gate-vertex set from scale-free graph discovered by both approaches are slightly smaller than that of Erd\"{o}s-R\'{e}nyi graphs.
This is true for the number of edges in gate graphs as well, when $\epsilon$ is no less than $3$.

In general, the running time of both approaches on Erd\"{o}s-R\'{e}nyi graphs are faster than that of scale-free graphs.
Especially, since for relatively large $\epsilon$, the size of gate-vertex set discovered by SC is significantly smaller than that of FS, we also observe that FS takes longer time than SC on those datasets with large value of $\epsilon$.

\comment{
When $\epsilon$ is $3$, the running time of SC on graphs with various edge density are $5$s, $28$s, $93$s, $247$s and $548$s;
and the running time of FS are $92$ms, $152$ms, $230$ms, $326$ms and $450$ms, respectively.
}

\comment{
In this experiment, we generated a set of random graph based on Erd\"{o}s-R\'{e}nyi model, with the edge density from
$2$ to $6$, while keeping the number of vertices at $10K$.
Here, we also vary the locality parameter $\epsilon$ from $2$ to $6$ to investigate the size compression of simplified graph compared to original ones.
Table \ref{tab:randomresult} shows original graph diameter, average pairwise shortest distance, the number of vertices and the number of edges in the simplified graph.
The distance distribution on those graphs is reported in Figure~\ref{fig:rand10kdist}.
The observations from scale-free graph are still hold on Erd\"{o}s-R\'{e}nyi random graphs.
Given the same $\epsilon$ and edge density, the number of gate-vertex from scale-free graph is slightly smaller than that of Erd\"{o}s-R\'{e}nyi graphs.
Interestingly, in terms of the number of edges in gate graphs, scale-free graphs achieve much better results than the one of Erd\"{o}s-R\'{e}nyi graphs, by an average of $16\%$, $20\%$, $30\%$ reduction on $\epsilon=2$, $\epsilon=3$ and $\epsilon=4$, respectively.

The running time of Erd\"{o}s-R\'{e}nyi graphs is faster than that of scale-free graphs.
When locality parameter is specified to be $2$, their running time for various edge density are $5$s, $30$s, $106$s, $260$s, and $614$s, respectively.
As $\epsilon$ is set to be $5$, the running time grows to be $402$s, $1810$s, $2159$s, $2612$s, and $3510$s, respectively.
}

\comment{
As large amount of shortest paths are taken as local distance, we can see the number of gates dramatically drops when the locality parameter goes beyond average shortest distance (i.e., the peek in Figure~\ref{fig:rand10kdist}).
For decreasing trend of the number gates, we have following observations:

\noindent 1) when locality parameter is smaller than average shortest distance ({\bf Avg.Dist}), as $\epsilon$ increases, more shortest paths are needed to be covered by gate vertices.
In addition, longer shortest paths share more common vertices which can be chosen as gate.
Therefore, only smaller number of gates are needed even more longer shortest paths have to be considered.

\noindent 2) when locality parameter just goes beyond average shortest distance ({\bf Avg.Dist}), as $\epsilon$ increases, fewer shortest paths are needed to be covered by gate vertices.
Combining the effect of common vertices shared by such long shortest paths, reduction of the number vertices in the simplified graph is even enforced.
Thus, the significant dropping of the number of discovered gates is observed.

The number of edges in the simplified graph is significantly affected by the number of discovered gates.
In other words, it decreases only when the number of gates is dramatically reduced because most shortest paths are considered as local paths.
}


\noindent{\bf Large Random Graph: }
Finally, we perform this experiment on a set of Erd\"{o}s-R\'{e}nyi random graphs and scale-free random graphs with average edge density of 2, and we vary the number of vertices from $100K$ to $500K$.
The locality parameter $\epsilon$ is specified to be $4$.
The number of vertices and the number of edges in gate graphs are shown in Figure~\ref{fig:ggvertexrandom} and Figure~\ref{fig:ggedgerandom}, respectively.
The diameter of 5 Erd\"{o}s-R\'{e}nyi graphs are $18$, $18$, $19$, $19$, $21$, and their average values of pairwise distance are $8.4$, $8.9$, $9.2$, $9.4$ and $9.6$.
For $5$ scale-free graphs, their diameter are $12$, $13$, $13$, $14$, $15$, and average values of pairwise distance are $7.8$, $8.2$, $8.4$, $8.6$ and $8.7$.

As we can see, the size of gate-vertex set discovered by SC is consistently smaller than that of FS in both types of graphs,
while FS outperforms SC in terms of the number of edges in gate graphs.
Moreover, as the number of vertices in original graphs increases, the size of gate-vertex set discovered by SC grows slower than that of FS.
For both FS and SC, the number of discovered gate vertices from scale-free graphs are smaller than the one from Erd\"{o}s-R\'{e}nyi graphs.

Overall, we observe that the reduction ratio of gate vertices on the real-world graphs is significantly better than that of the synthetic graphs.
This suggests that in the real world graphs, its underlying structure is not that ``random''.
In other words, the real graphs seems to have more recognizable ``highway'' structure in terms of the shortest path connection.
From this perspective, the existing research on random graph generators have not been able to model this network behavior.

\comment{
As we can see, when $\epsilon=2$, the number of gate-vertex discovered from Erd\"{o}s-R\'{e}nyi is slightly smaller than that of scale-free graphs.
However, when $\epsilon$ is greater than $2$, scale-free is the winner in terms of both the number of gate-vertex.
More significantly, as locality parameter $\epsilon$ increases, the number of edges generated from scale-free graphs grows much slower than the one from Erd\"{o}s-R\'{e}nyi graphs.
This is because each discovered gate-vertex in scale-free graphs tends to cover more vertex pairs within gate graph than Erd\"{o}s-R\'{e}nyi graphs.
}

\section{Conclusion}
\label{conclusion}

In this paper, we study a new graph simplification problem to provide a high-level topological view of the original graph while preserving distances.
Specifically, we develop an efficient algorithm utilizing recursive nature of shortest paths and set cover framework to discover gate-vertex set.
More interestingly, our theoretical results and algorithmic solution can be naturally applied for minimum $k$-skip cover problem, which is still open problem.
In the future, we would like to study whether approximate distance with guaranteed accuracy can be gained based on our framework.
We also want to investigate how our simplified graph can be applied for graph clustering, multidimensional scaling and graph visualization.

\bibliographystyle{plain}
\bibliography{bib/ComplexNetwork,bib/ComplexNetwork2,bib/simplification,bib/uncertaingraph,bib/cikm,bib/reachability,bib/3hop,bib/reachpaper,bib/Yangdissertation,bib/distance,bib/distance10}

\end{document}